\documentclass[useAMS,usenatbib]{mnras}
\usepackage{amssymb}
\usepackage{graphicx}
\usepackage{caption}
\usepackage{subcaption}
\usepackage{amsmath}
\usepackage{epsfig}
\usepackage{float}
\usepackage{rotating}
\usepackage{url}
\usepackage{epstopdf}
\usepackage{mathabx}
\usepackage{fnpos}
\usepackage{longtable} 
\usepackage{lscape}
\usepackage{multirow}
\usepackage{rotating}
\usepackage[usenames,dvipsnames]{color}
\usepackage{mathtools}
\DeclarePairedDelimiter\abs{\lvert}{\rvert}%
\DeclarePairedDelimiter\norm{\lVert}{\rVert}%

\makeatletter
\let\oldabs\abs
\def\abs{\@ifstar{\oldabs}{\oldabs*}}
\let\oldnorm\norm
\def\norm{\@ifstar{\oldnorm}{\oldnorm*}}
\makeatother
\newcommand{\ssun}{{$_{\Sun}$}}	

\newcommand{\HI}{H \textsc{i}}
\newcommand{\HII}{H \textsc{ii}}

\newcommand{\gre}{\textgreater} 
\newcommand{\les}{\textless}
\newcommand{\apr}{$\sim$}
\newcommand{\x}{$\times$}
\newcommand{\degrr}{$^\degree$}
\newcommand{\e}[1]{10$^{#1}$}
\newcommand{\Vhb}{$V_{\mathrm{HB}}$}

\newcommand{\per}{$^{-1}$}
\newcommand{\mean}[1]{$\langle$#1$\rangle$}

\title[Ca II triplet spectroscopy of RGB stars in NGC~6822]{Ca II triplet spectroscopy of RGB stars in NGC~6822: kinematics and metallicities}
\author[J. Swan, A. A. Cole, E. Tolstoy, M. J. Irwin.]{J. Swan$^{1}$\thanks{E-mail: jesse.swan@utas.edu.au}, A. A. Cole$^{1}$, E. Tolstoy$^{2}$, and M. J. Irwin$^{3}$.  \\
$^{1}$School of Physical Sciences, University of Tasmania, Private Bag 37, Hobart, Tasmania, 7001 Australia\\
$^{2}$Kapteyn Astronomical Institute, Landleven 12, NL-9747 AD Groningen, the Netherlands\\
$^{3}$Institute of Astronomy, University of Cambridge, Madingley Road, Cambridge CB3 0HA, UK\\}

\begin{document}

\date{Accepted YYYY Month DD. Received YYYY Month DD; in original form YYYY Month DD}

\pagerange{\pageref{firstpage}--\pageref{lastpage}} \pubyear{PUBYEAR (YYYY)}

\maketitle

\label{firstpage}

\begin{abstract}
We present a detailed analysis of the chemistry and kinematics of red giants in the dwarf irregular galaxy NGC~6822. Spectroscopy at $\approx\,8500$ \AA{} was acquired for 72 red giant stars across two fields using FORS2 at the VLT. Line of sight extinction was individually estimated for each target star to accommodate the variable reddening across NGC~6822. The mean radial velocity was found to be \mean{$v_{rad}$} $=-52.8\pm2.2$ km s\per{} with dispersion $\sigma_{v}=24.1$ km s\per{}, in agreement with other studies. Ca \textsc{ii} triplet equivalent widths were converted into [Fe/H] metallicities using a $V$ magnitude proxy for surface gravity. The average metallicity was \mean{[Fe/H]} $=-0.84\pm0.04$ with dispersion $\sigma=0.31$ dex and interquartile range 0.48. Our assignment of individual reddening values makes our analysis more sensitive to spatial variations in metallicity than previous studies. We divide our sample into metal rich and metal poor stars; the former were found to cluster towards small radii with the metal poor stars more evenly distributed across the galaxy. The velocity dispersion of the metal poor stars was found to be higher than that of the metal rich stars ($\sigma_{v_{\rm MP}}=27.4$ km s\per{}; $\sigma_{v_{\rm MR}}=21.1$ km s\per{}); combined with the age-metallicity relation this indicates that the older populations have either been dynamically heated during their lifetimes or were born in a less disklike distribution than the younger stars.. The low ratio $v_{rot}/\sigma_v$ suggests that within the inner 10$\,'$, NGC~6822's stars are  dynamically decoupled from the H I gas, and possibly distributed in a thick disc or spheroid structure. 
\end{abstract}


\section{Introduction}
Dwarf galaxies are by far the most numerous galaxy class in the Universe; the VIMOS Public Extragalactic Redshift Survey (VIPERS) estimates that dwarf galaxies account for \apr{}95\% of all galaxies in the visible Universe \citep{Bel2013}. Hierarchical structure formation of galaxies invokes a population of early protogalaxies, similar in mass to present-day dwarf galaxies,  from which present day large galaxies formed (e.g., \citealt{Lacey1993}). Whilst these proto-galactic fragments aren't exactly the same as today's dwarf population, study of the latter can still grant insight into galaxy assembly, interactions, gas cooling, and low-metallicity star formation in shallow gravitational potential wells. In particular, the red giants in dwarf galaxies of the Local Group can be observed with a high enough spectral resolution and signal-to-noise (S/N) that we can obtain accurate radial velocities and metallicities of large samples of stars. We do not have this luxury with larger elliptical and spiral galaxies simply because of their scarcity within distances of $\sim1$ Mpc.

NGC~6822 is a historical icon due to the Cepheid variable observations by \cite{Hubble1925} placing it as the first indisputably extragalactic object. Current measurements put NGC~6822 at $488\pm5$ kpc \citep{Feast2012} making this galaxy among the closest non-satellites to the Milky Way (MW). NGC~6822 is a fairly typical dwarf irregular (dI) having a radius \apr{}2.6 kpc, \HI{} mass of \apr{}1.64\x\e{9} M\ssun{} \citep{DeBlok2000}, and luminosity $M_{B}\sim -15.8$. UV star formation (SF) estimates put the star formation rate (SFR) of NGC~6822 over the last 100 Myr to be $\sim0.014$ M\ssun{} yr\per{} \citep{Efremova2011}. NGC~6822 lies near the galactic plane ($l,b=25.3$\degrr,-18.4\degrr) which results in moderate foreground extinction: $0.24\leq{}E(B-V)\leq{}0.37$ (\citealt{Schlegel1998}; \citealt{Fusco2012}). 

NGC~6822 is the closest of $\approx$8--10 galaxies of similar stellar and total mass in and around the Local Group, and as such is a natural target for study of its chemical evolution and star formation history. However, it also stands out as an object of particular interest due to its disturbed HI morphology (\citealt{Skillman1989},~\citealt{DeBlok2000}) and the differentiated spatial distribution of its stellar populations ({ e.g., \citealt{Hodge1991}, \citealt{DeBlok2006}, \citealt{Sibbons2012}}), which give it the potential to aid the understanding of the progression of galaxy interactions in the regime M$_{dyn}$ $\lesssim$10$^9$ M$_{\odot}$.

The young stellar populations and star clusters of NGC~6822 have been historically studied by a number of authors as summarized by \cite{Hodge1977} and by \cite{Hoessel1986}. The first large-scale effort to establish a quantitative star-formation history was made by Gallart et al.\ (1996a,b,c), who found evidence that the sheet of red stars pervading NGC~6822 is at least several billion years old, and attempted to establish a date for the epoch of first star formation in NGC~6822. This attempt was limited by the lack of knowledge of the chemical evolution of the galaxy, owing to the relatively shallow photometric limit of ground-based data and the age-metallicity degeneracy for red giants.  Gallart et al.\ (1996b) concluded that the star formation rate in NGC~6822 has been declining over the past $\sim$3~Gyr. 

The first HST observations of NGC~6822
\citep{Wyder2001} took advantage of the improved spatial resolution to refine the view of the intermediate-age populations, and came to the opposite conclusion, finding that over the past few gigayears NGC~6822 has been in a period of increased star-forming activity relative to its lifetime average. This result was confirmed and extended beyond the central bar region by subsequent colour-magnitude diagram analyses \citep[e.g.,][]{Cannon2012}, supported by spectroscopic metallicity measurements of star cluster integrated light \citet{Cohen1998} and HII regions \cite{Skillman1989a}. 

NGC~6822 has remained an object of intense interest in recent years, as investigators have studied the radial gradients and connections between the stars and gas in attempts to understand the disturbed gas morphology and the nature of the extended distribution of the field stars (e.g., Letarte et al.\ 2002, Battinelli et al.\ 2003, Fusco et al.\ 2014) and clusters (e.g., Hwang et al.\ 2014). A wealth of new photometric and spectroscopic data has led to sometimes conflicting claims about the structure and kinematics of the old and intermediate age populations (e.g., Demers et al.\ 2005) with respect to the young stars and gas (e.g., de Blok \& Walter 2006).

In this paper we present a detailed analysis of 72 RGB stars in NGC~6822 drawn from the southern part of the central bar and from a field several arcminutes to the east. Calcium II triplet (CaT) spectroscopy is used to determine radial velocities and [Fe/H]\footnote{[Fe/H]=log(Fe/H)$_\star$-log(Fe/H)\ssun} iron abundances of the intermediate-age and old stellar population. These results are able to be compared directly to the first RGB abundances and stellar velocities determined by \cite{Tolstoy2001} and the more extensive abundances from \cite{Kirby2013}, along with the detailed stellar velocities from \cite{Kirby2014}. The observations and data reduction are presented in section 2 and in section 3, we address the issue of differential reddening from dust in the Milky Way and in NGC~6822. In section 4 we calculate the equivalent widths of calcium, magnesium, and sodium lines in the near infrared in order to identify foreground Milky Way stars that overlap in velocity with NGC~6822, and to calculate the metallicity of NGC~6822 members. In section 7 we analyse the resulting radial velocity and metallicity distributions to search for radial gradients, kinematic-metallicity correlations indicative of multiple stellar populations, and signs of rotational support. 

\section{Data acquisition and processing}
The observations for this dataset were made over several nights in late July and early August of 2003 using the FORS2 instrument mounted on the Yepun UT4 8m telescope of the ESO VLT in Chile. Pre-imaging for relative photometry and astrometry was obtained prior to the observing run and used to select targets for spectroscopic study. The complete log of observations is given in Table~\ref{tab:obslog}. 

\begin{table*}
  \centering
  \caption[Observing Log]{Pre-Imaging and Spectroscopic Observations of NGC~6822 Stars}
    \begin{tabular}{lcccccc}
    \hline
    \hline
      Field & Position  & Date/Time & Spectral & Exp & Airmass & Seeing \\
      & ($\alpha$, $\delta$, J2000.0)   & (UT) & Element & (s)   &  & ($^{\prime\prime}$) \\
    \hline
     \multicolumn{7}{c}{Pre-Imaging}\\
      NGC6822-E  & 19:45:34, $-$14:49:50 & 2003-04-14/08:49 & Bessell V &  60  & 1.18 & 0.59 \\
      NGC6822-E  &                                     & 2003-04-14/08:51 & Bessell I &  60  &  1.17 & 0.65 \\
      NGC6822-S  & 19:44:58, $-$14:52:23 & 2003-08-22/04:30 & Bessell V & 45 & 1.16 & 0.61 \\
      NGC6822-S  &                                     & 2003-08-22/04:32 & Bessell I & 45 & 1.17 & 0.64 \\
    \hline
     \multicolumn{7}{c}{Slitmask Spectroscopy}\\
      NGC6822-E  & 19:45:34, $-$14:49:50 & 2003-08-22/00:01 & 1028z grism & 2400 & 1.23 & 0.49 \\
      NGC6822-E  &                                     & 2003-08-22/00:43 &                      & 2400 & 1.12 & 0.71 \\
      NGC6822-E  &                                     & 2003-08-22/01:42 &                      & 2400 & 1.03 & 0.62 \\
      NGC6822-E  &                                     & 2003-08-22/02:23 &                      & 2400 & 1.01 & 0.56 \\
      NGC6822-E  &                                     & 2003-08-22/03:06 &                      & 2400 & 1.03 & 0.51 \\
      NGC6822-E  &                                     & 2003-08-22/03:48 &                      & 2400 & 1.07 & 0.53 \\
      NGC6822-S  & 19:44:58, $-$14:52:23 & 2003-08-22/23:52 & 1028z grism & 2400 & 1.25 & 0.73 \\
      NGC6822-S  &                                     & 2003-08-23/00:34 &                      & 2400 & 1.13 & 0.70 \\ 
      NGC6822-S  &                                     & 2003-08-23/01:28 &                      & 700$^{\dagger}$  & 1.04 & 0.85  \\ 
      NGC6822-S  &                                     & 2003-08-23/23:41 &                      & 2400 & 1.28  & 1.00 \\ 
      NGC6822-S  &                                     & 2003-08-24/00:24 &                      & 2400 & 1.14  & 1.10 \\ 
      NGC6822-S  &                                     & 2003-08-24/01:08 &                      & 2400 & 1.06  & 0.90 \\ 
      NGC6822-S  &                                     & 2003-08-24/02:03 &                      & 2400 & 1.02  & 0.73 \\ 
      NGC6822-S  &                                     & 2003-08-24/02:51 &                      & 2400 & 1.02  & 0.88 \\ 
      NGC6822-S  &                                     & 2003-08-24/03:33 &                      & 2400 & 1.06  & 0.68 \\ 
    \hline
    \end{tabular}\\
    $^{\dagger}$Exposure terminated due to cloud; not used in analysis.
  \label{tab:obslog}%
\end{table*}%

\subsection{Target selection}
\begin{figure*}
  \centering
  \includegraphics[width=0.97\textwidth]{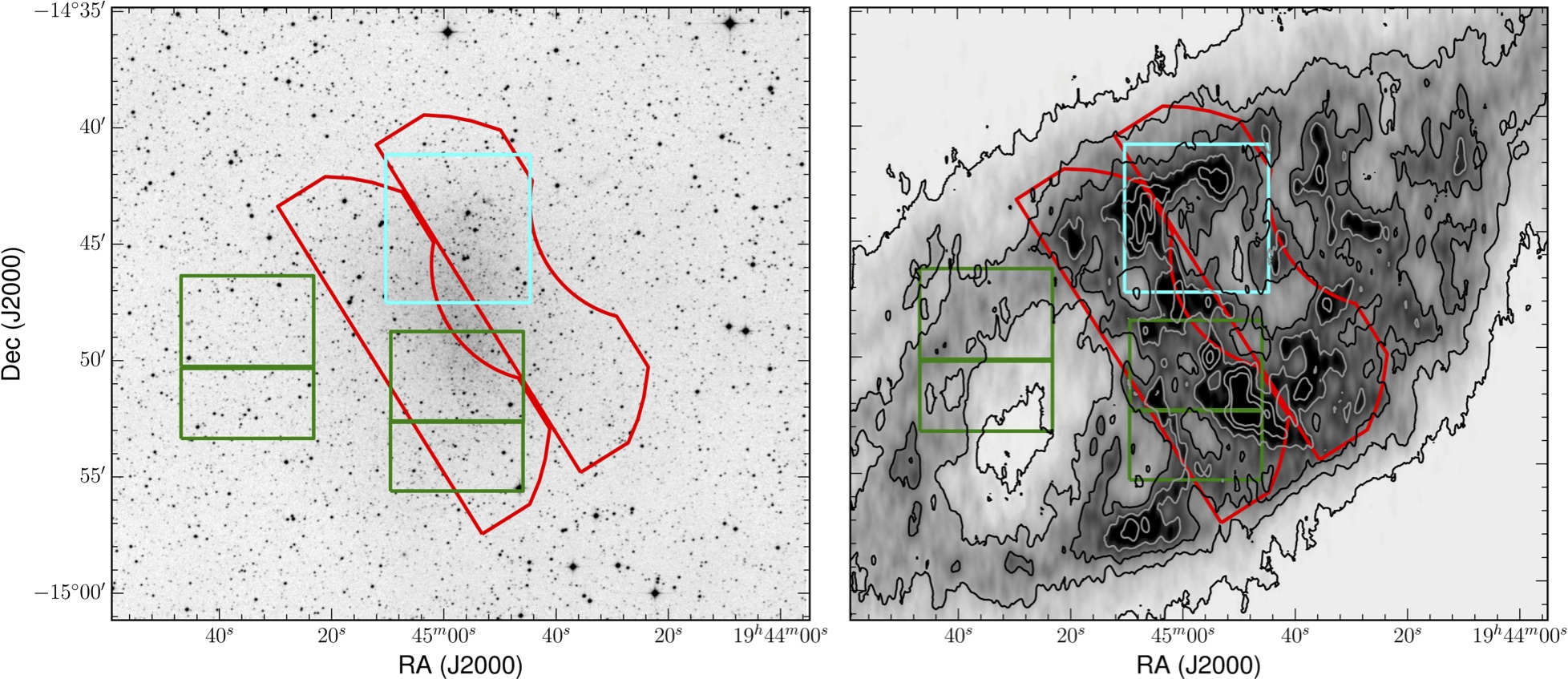}
  \caption[NGC~6822 image and H\textsc{i} contours]{Each of these frames represents a $30'\times{}30'$ view centred on NGC~6822. Frame (a) is imagery acquired with the Anglo-Australian telescope obtained from the Digital sky survey\protect\footnotemark, and frame (b) is a contour plot of integrated \HI{} column density of the same region \citep{DeBlok2000}. Contour levels are 1,7,13,\ldots31\x\e{20} cm$^{-1}$. North is up, and East is left. The South and East field surveyed in this work are indicated with rectangular boxes. The left-most pair of rectangles in each frame are representative of the total field of view (FoV) of the East exposures, the right-most are representative of the South region. Each of the top larger rectangles correspond to the master chip FoV, the bottom is the slave FoV. The two large red regions are the Keck/DEIMOS fields observed in Kirby et al. (2013), and the cyan square is the region covered by Tolstoy et al. (2001).}
  \label{fig:CCDregs}
\end{figure*}

Our targets were chosen from two regions in NGC~6822 which were selected to correspond to the galaxy's high density central region (hereafter South region) and a much lower density outer region of the galaxy (hereafter East region; see Figure \ref{fig:CCDregs}). Within these regions, targets were selected using 60 second FORS2 pre-imaging in \textit{V}- and \textit{i}-band. Individual stars were then chosen from this photometry to have colours consistent with RGB membership and to be within half a magnitude of the tip of the red giant branch (TRGB). While this maximizes the probability that our sample will consist of intermediate-age and old first ascent red giants, contamination by low-luminosity red supergiants and early asymptotic giants remains a possibility. In the selection phase we were unable to eliminate contamination of our sample by foreground MW dwarf stars, which is of particular importance due to the location of NGC~6822 near the galactic plane.

\footnotetext{Data from this survey hosted at the Canadian Astrophysics Data Centre CADC - \\ \url{http://www3.cadc-ccda.hia-iha.nrc-cnrc.gc.ca/community/STETSON/standards/}}

\subsection{Observations} % (fold)
\label{sub:observations}
Spectral observations were conducted using the Mask eXchange Unit (MXU) mode of the FORS2 instrument mounted on the UT4 Cassegrain focus \citep{Appenzeller1998}. Using the MXU our 94 targets were observed through $1''\times8''$ slits ($1''\times7''$ in a few cases to prevent spectral overlap) based on accurate relative astrometry from the FORS2 pre-images. This choice of slit length, whilst conservative, guarantees a localized background level for subtraction from each stellar spectrum. Furthermore, it easily allows successive observations to be { dithered} by \apr{}2$''$ along their 8$''$ lengths in order to reduce complications from bad pixels, cosmic rays, and sky fringing.

The spectra were obtained at a plate scale of 0.25$''$ pixel\per{}, and a 6.8$'$\x{}6.8$'$ field of view granted via use of the Standard Resolution Collimator. This was used in conjunction with the 1028z+29 holographic grism and OG590$+$32 order blocking filter to grant a spectral dispersion of \apr{}0.85 \AA{} pixel\per{} over $\lambda=7700$ \AA{} to $9500$ \AA{} { (resolving power R $\approx$3400)}. The spectra were recorded on the MIT/LL 2k\x{}1k CCD (15$\mu$ pixel width) with a gain of 0.7 ADUs/e$^-$; the `slave' and `master' CCD had read noise 2.7e$^-$, and 3.15e$^-$ respectively. In order to accommodate the spectral range of interest (the Ca \textsc{ii} triplet; $\lambda$ $\approx$ 8490 to 8670 \AA{}) the effective field of view of the mosaic was limited to 6.8$'$\x{}5.7$'$. The observing log is shown in Table~1; the South field had an exposure time of 8$\times$2400~s with seeing between 0.63$''\leq$ FWHM $\leq 1.22''$, and the East field had a total integration time of 6$\times$2400~s with 0.44$''\leq$ FWHM $\leq 0.71''$.

The data reduction process was completed via primary use of native IRAF tasks in conjunction with several custom IRAF scripts. The task \textsc{ccdproc} was used on each of the exposures and their appropriate bias and flat field master frames to routinely apply a bad-pixel mask to the science images, subtract the over scan regions and trim the images whilst simultaneously applying the bias correction and flat fielding. 
% subsection observations (end)

\subsection{Distortion correction}
Distortions in the raw spectra necessitated that geometric corrections be made prior to extraction and analysis. These corrections were made in a two-step process, each process using a different custom IRAF script (see \citealt{Leaman2009}). The first script traced each bright stellar continuum and applied an appropriate quadratic correction about the CaT. The second script then traced each sky line and applied a suitable linear correction to orthogonalize the skyline to the stellar continuum. This correction not only reduces the complexity in spectral extraction and dispersion solution (and in turn the radial velocity measurements), but also improves the accuracy of the sky background subtraction.

\subsection{Wavelength dispersion solution}
\begin{figure}
  \centering
  \includegraphics[width=0.48\textwidth]{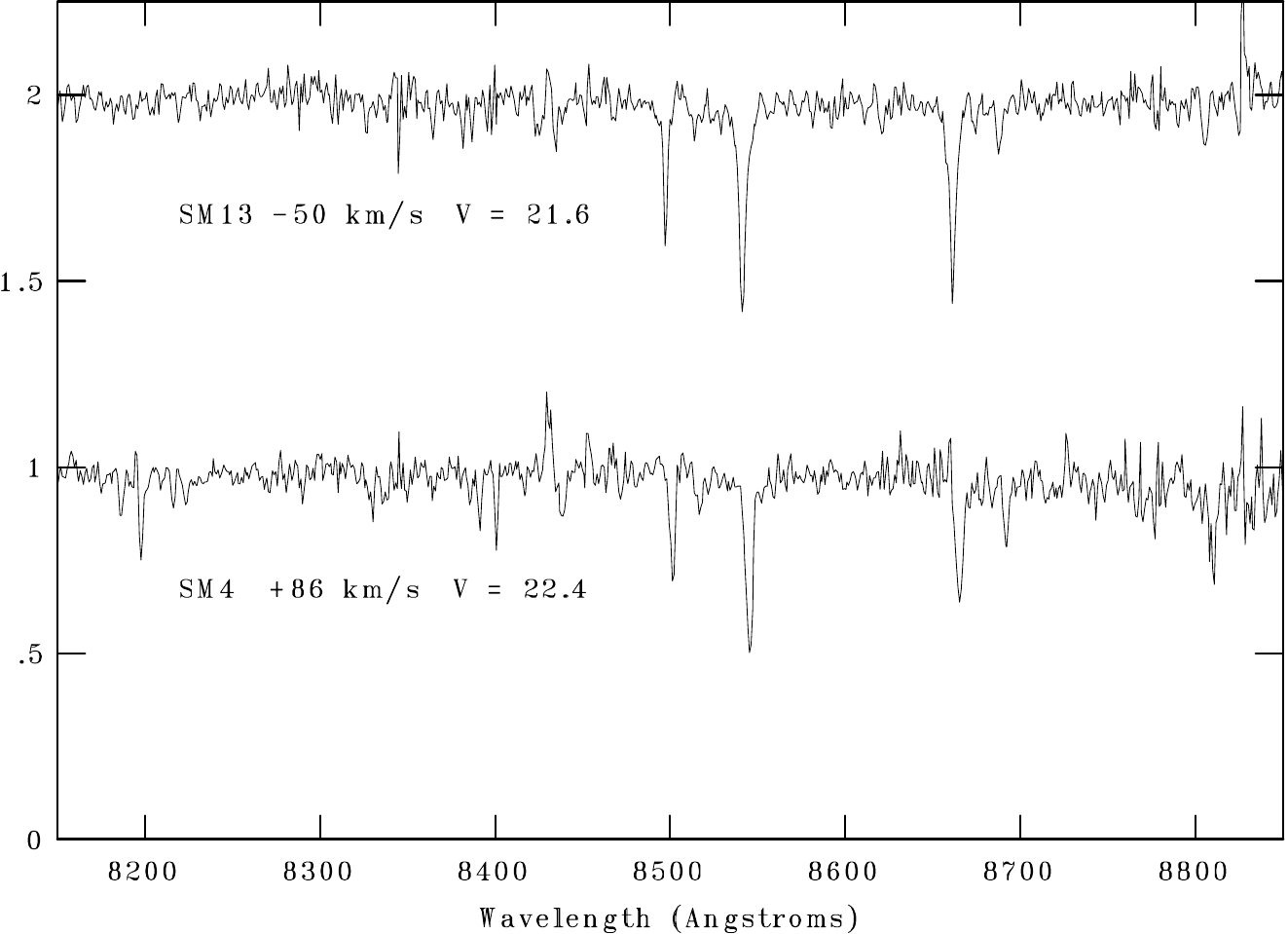}
  \caption[]{The final stacked spectra of two of our target stars; velocity corrected and three pixel smoothed. The star ID, radial velocity, and V-band magnitude are listed beneath each spectrum. The upper spectrum is that of a red giant star in NGC~6822. The lower spectrum is a foreground Milky Way dwarf contaminant. Note the dominant Na \textsc{i} doublet ($\lambda = 8183,8195$ \AA{}) in the dwarf spectrum that is not present in the red giant spectrum. The Mg \textsc{i}  ($\lambda = 8807$ \AA{}) line is also more dominant in the dwarf star.}
  \label{fig:spectra}
\end{figure}

The wavelength dispersion solution was generated using between 35 and 50 OH, and O$_2$ emission lines from the spectra. Wavelengths of these emission lines were taken from the \cite{Osterbrock1996} night sky atlas. The resultant RMS accuracy was between \apr{}0.05 and 0.10 \AA{}. Night sky emission lines were used for applying the wavelength dispersion rather than daytime arc lamp exposures as this alleviates the effect of any possible drift due to temperature variations or spectroscopic flexure \citep{Cole2004}. The individual stellar spectra were then extracted using the \textsc{hydra} package within IRAF and the sky subtraction performed. The final clean spectra for each of the 94 stars were then { normalised using the IRAF \textsc{continuum} task by fitting a fifth-order cubic spline to the continuum, excluding a 100 pixel region on either end of the spectrum to prevent numerical instability,} and combined with \textsc{scombine} to boost the S/N of the final spectra. {  Figure \ref{fig:spectra} is an example of a RGB member of NGC~6822 and a star that was identified as a foreground Milky Way dwarf contaminant to our sample.}

\subsection{Radial velocity measurements}
The radial velocities of our target stars were measured by way of a Fourier cross-correlation between our spectra and 24 calibrated template spectra of Milky Way RGB\footnote{One of the template stars is a dwarf star; whilst this does not alter the velocity calculations it is exploited later on.} members of well-studied star clusters (see Table \ref{tab:templates}). The Fourier cross-correlation \citep{Tonry1979} was performed using the \textsc{fxcor} task in IRAF.  The typical { random} uncertainty in the velocity measurements was identical across both the South and East field, having a median absolute uncertainty $\langle\delta v_{rad}\rangle$ = 1.28 km s$^{-1}$. Slit centering corrections needed to be applied due to systematic offsets in the individual slits. Given the spectral resolution and CCD pixel scale, slit-centring errors can contribute up to 29.5 km s$^{-1}$ pixel$^{-1}$.  { We calculated the required corrections by examination of through-slit images taken without the grating; the stellar centroid position in the dispersion direction was compared to the centroids of the slits themselves in order to determine the correction.} Typical corrections were on the order of $\Delta V_{rad}$ \apr 0.5 - 1.2 km s$^{-1}$. { The velocity error budget is dominated by the uncertainty in the slit-centring correction, and contains a contribution from the dispersion of velocities derived from the 24 individual comparison template spectra as well; typical final errors are on the order of 6--7~km~s$^{-1}$.}

The spectrum of the lower-most slit on the East master frame was discarded due to contamination via scattered light from the MXU on FORS2. A thorough visual inspection of the spectra of the remaining 93 stars revealed two carbon stars in our sample (one from each the South and East field). Despite the strong CN and C$_2$ molecular bands the CaT is still present so velocities were still obtained for these stars (see Table \ref{tab:bigtableofstardata}). The relation between the CaT and the continuum however, is affected to the point that determination of the equivalent widths of these spectral lines would no longer be useful. Thus, no metallicity analysis was performed on these stars. 

\section{Equivalent width measurements} % (fold)
\label{sec:equivalent_width_measurements}
The equivalent width measurements were performed using a modified version of interactive software written and kindly provided by \cite{DaCosta1999}
to A. Cole; this program was modified to fit the spectral lines using the sum of a Lorentzian and a Gaussian profile \citep{Cole2004}, measured relative to a linear fit to the continuum bandpasses. The bandpasses are those defined by \citet{Armandroff1988}. The median fractional uncertainty for these measurements were: $\langle \delta W\rangle/W= 0.029$, 0.044, and 0.079 - ordered from the strongest to weakest of the CaT lines. The benefit of using the sum of a Lorentzian and a Gaussian function is outlined in \cite{Cole2004}, but arguably is only really of importance for spectra with a higher S/N than our own. { We experimented with different continuum-fitting approaches and with direct numerical integration vs.\ profile-fitting and found that no reasonable change produced a large change in the measured equivalent width, amounting to a total plausible systematic error of $\lesssim$0.09~\AA. }The same program was used to measure neutral sodium and magnesium lines in the nearby spectrum during the identification of foreground contaminants (see below).
% section equivalent_width_measurements (end)

\section{Calculation of [Fe/H]} % (fold)
\label{sec:calculation_of_met}
Since the CaT was first identified as a good metallicity indicator \citep{Armandroff1988} several studies have sought to explore its robustness over wide ranges of stellar ages and metallicities (e.g., \citealt{Rutledge1997}; \citealt{Cole2004}; \citealt{Carrera2007}, \citealt{Starkenburg2010}, \citealt{Carrera2013}, \citealt{Moto'oka2013}). In this work, we adopt the empirical calibrations derived in \cite{Cole2004}, calibrations that used the same FORS2 instrumentation as this study and very similar data reduction techniques. Whilst all calibrations utilize the summed equivalent widths of the CaT in either a weighted (e.g., \citealt{Rutledge1997}) or unweighted fashion (e.g., \citealt{Olszewski1991}; \citealt{Cole2004}; \citealt{Leaman2009}), some authors choose to omit the weakest of the calcium triplet members ($\lambda=$ 8498 \AA{}) due to the potential reduction to S/N (e.g., \citealt{Suntzeff1993}; \citealt{Cole2000}) whilst other authors - in particular those with high S/N - use all three lines (e.g., \citealt{Cole2004}; \citealt{Leaman2009}). In this work, we accept the the additional noise and use an unweighted sum of the equivalent width of all three lines\footnote{For a full `three line justification' for data obtained in the same observing run as our own, see \cite{Leaman2009} \S 3.1.3} for each of our 91 RGB stars:
\begin{align} 
\Sigma W = W_{8498} +W_{8542} +W_{8662}. \label{eq:sumW}
\end{align}
The reduced equivalent width or `calcium index', $W'$ is then constructed in the manner outlined in \cite{Armandroff1991}:
\begin{align} 
W'=\Sigma W +\beta(V-V_{\mathrm{HB}}) \label{eq:Wprime}
\end{align}
This particular equivalent width measure exploits the empirically determined linear relation between an RGB stars $V$ magnitude and its measured $\Sigma W$ for a given stellar population (see e.g., Figure 3, \citealt{Armandroff1991}). Scaling the summed equivalent width by the magnitude of the target star adequately removes any temperature or gravity dependence of the Ca \textsc{ii} line widths by assuming that the function describing this dependence is linear over the small magnitude range $V-V_{\mathrm{HB}}$. It is this strong dependence of the calcium triplet line widths on surface gravity and effective temperature that prevents us from quantitatively analysing the metallicities of the MW foreground dwarfs identified in this sample, as the correction introduced in equation (\ref{eq:Wprime}) is not valid for stars outside of the RGB. 

The use of the $V-V_{\mathrm{HB}}$ term in equation (\ref{eq:Wprime}) rather than an absolute magnitude $M_V$, serves as  a correction for the distance and reddening to the target population; because distances are typically uncertain around the $\approx$10\% level, a relative measure is frequently preferable. However, differential reddening across the target field can result in variations such that use of a single $V-V_{\mathrm{HB}}$ value for a whole galaxy can introduce biases to the derived metallicities. Most authors take \Vhb{} to be an average reddening corrected value across the entire galaxy or cluster (e.g., \citealt{Armandroff1991}; \citealt{Cole2004}; \citealt{Leaman2009}), whilst this is perfectly valid in almost all cases, NGC~6822 has a relatively high and patchy reddening \citep[e.g.,][]{Fusco2012}. Therefore we explored a unique method of determining individually reddening-corrected synthetic \Vhb{} magnitudes for each of our stars using the variation in the tip of the red giant branch magnitude around our targets to determine the appropriate synthetic \Vhb\ to use in Equation~\ref{eq:Wprime} (see \S{} \ref{sec:redd}). The parameter $\beta$ is ideally independent of metallicity over a wide metallicity range so as to justify the linearity of equation (\ref{eq:Wprime}). Here we will use $\beta = 0.73 \pm 0.04$ from \cite{Cole2004} as this was uniquely derived to correspond to the same Lorentzian~Gaussian fit to the spectral lines as was utilized in this paper.

\subsection{Reddening correction} \label{sec:redd}
Differential reddening in front of and within NGC~6822 causes \Vhb\ to vary from region to region, which may cause stars with identical atmospheric parameters to appear to vary in metallicity depending on their location in the galaxy. Wholly accounting for the reddening of NGC~6822 along each line of sight is a non-trivial exercise. Both the internal reddening due to the ISM of NGC~6822 and the extinction along our line of sight in the MW combine to give reddening values higher than encountered in spectroscopic studies of similar galaxies (e.g., \citealt{Leaman2009}; \citealt{Parisi2009}; \citealt{Foster2010}; \citealt{Depropris2010}). The Schlegel, Finkbeiner, \& Davis \citeyearpar{Schlegel1998} dust maps give a foreground extinction of $E(B-V)=0.24$ based on dust column density surveys; however, the reddening map derived by Battinelli, Demers \& Kunkel (2006) from these data show some hint of the NGC~6822 HI structure, indicating that at least some of the spatially variable dust emission originates in NGC~6822, where the conversion from far-infrared surface brightness to visual extinction may differ from that in the Milky Way. A recent HST study of the stellar populations by \cite{Fusco2012} gives a central $E(B-V)_C=0.37\pm0.02\,$, and an exterior $E(B-V)_E=0.30\pm0.03$ - these correspond roughly to our South and East fields, respectively. 

The significant variation between these two regions  can be briefly explored by considering the non-uniformity of the \HI{} gas within NGC~6822 (Figure \ref{fig:CCDregs}). Compared to the South field, the reddening due to internal \HI{} gas in the East region could be acceptably treated as spatially constant for the purpose of this study. However, because the \HI{} column density varies from roughly 4--31 \x\e{20} cm$^{-2}$ in the South region \citep{DeBlok2000}, and reddening scales roughly linearly with \HI{} column density, we cannot make the same assumption without compromising our determination of this region's metallicity. We therefore use an individual determination of the line of sight reddening to each of our 93 stars in order to reduce the systematic errors in final metallicity estimates; we rely on the relative reddening with respect to a known reference point. 

It is possible to determine the relative extinction between two regions by measuring the shift in magnitude of the tip of the red giant branch (TRGB) with respect to a reference region if the stellar populations are not too dissimilar. This has two advantages compared to attempting to directly use the variation in $V_{HB}$: the horizontal branch magnitude is itself a function of both metallicity and age, which may vary between fields independently of dust extinction (and which variation we do {\it not} wish to remove from the data); and calibrated, homogeneous photometry down to the level of the horizontal branch is not readily available for the entirety of both our fields. By contrast, the I-band absolute magnitude of the TRGB is a robust quantity that is almost insensitive to age over many gigayears, only mildly metallicity dependent, and precisely measurable across the full extent of NGC~6822. This will be our standard candle for determining the dereddening correction needed to synthesize a uniform V$_{HB}$ for use in Equation~\ref{eq:Wprime}.

Target positions from our FORS2 preimaging were cross-listed with 2.5m Isaac Newton Telescope (INT) $g$-, $r$-, and $i$-band photometry (\citealt{McMahon2001}) to obtain magnitudes for our targets in these bandpasses. The photometric $g$ and $i$ magnitudes were converted to $I$-band using transformations described in \cite{Windhorst1991}. Individual subregions were then constructed for each of our targets by selecting all objects in the \cite{McMahon2001} catalogue that fell within a 20$''$ region and had colours $i<22$ and $1.0<g-i<3.0$ to confine the photometry to the RGB. These subregions typically contain 9500 - 12500 objects for the East field targets and 14000 - 17000 for the South field target regions. 

Rather than manually determining the tip of the red giant branch (TRGB) location from CMDs, a one dimensional  Sobel edge-detection filter (e.g., \citealt{Sakai1996}) was applied to $i$-band magnitude distribution functions generated for each subregion. Since the TRGB location is the $i$-band magnitude of the maximum Sobel filter response, the final TRGB was taken as the average of this maximum from three Sobel edge-detections with slightly varied bandwidths (variation in TRGB location \les{}\apr{}0.02 mag); the random error was adopted as the standard deviation from the mean of the three measurements (see Table \ref{tab:errors}).

We desire the reference HB magnitude to correspond to a very well-measured, and, ideally, little-reddened region of NGC~6822. Conveniently, grid 5 of the HST photometry of \cite{Cannon2012} is centred on the giant HI hole (see Figure 1 in their paper), which also overlaps our East field. Transforming their F814W magnitudes (in the ACS photometric system) to standard Cousins I, the HB location of Cannon's grid 5 is found to be $I_{\mathrm{HB}}=23.09\pm0.05$. The final step is to convert this to a synthetic V$_{\rm HB}$ using a colour-estimate for the horizontal branch. High quality data for 47 Tucanae (which has a similar metallicity to NGC~6822; \citealt{Bergbusch2009}) and theoretical stellar isochrones consistently suggest that the intrinsic { average colour of the HB} is V$-$I $\approx$0.80.  { Each of these steps carries some associated uncertainty; combined with the intrinsic spread in colour and magnitude of the CMD features there is a likely total systematic error in the absolute V magnitude of the HB on the order of $\approx$0.2~mag.} The metallicity results are ultimately relatively insensitive to the exact value of V$_{HB}$. What we hope to have removed by this procedure is any small but systematic error that could lead to the spurious appearance or obscuration of a metallicity gradient in the final dataset. We comment on the possibility of the unintended introduction of a net shift in metallicity in the discussion below.

\begin{figure}
  \centering
\includegraphics[width=0.45\textwidth]{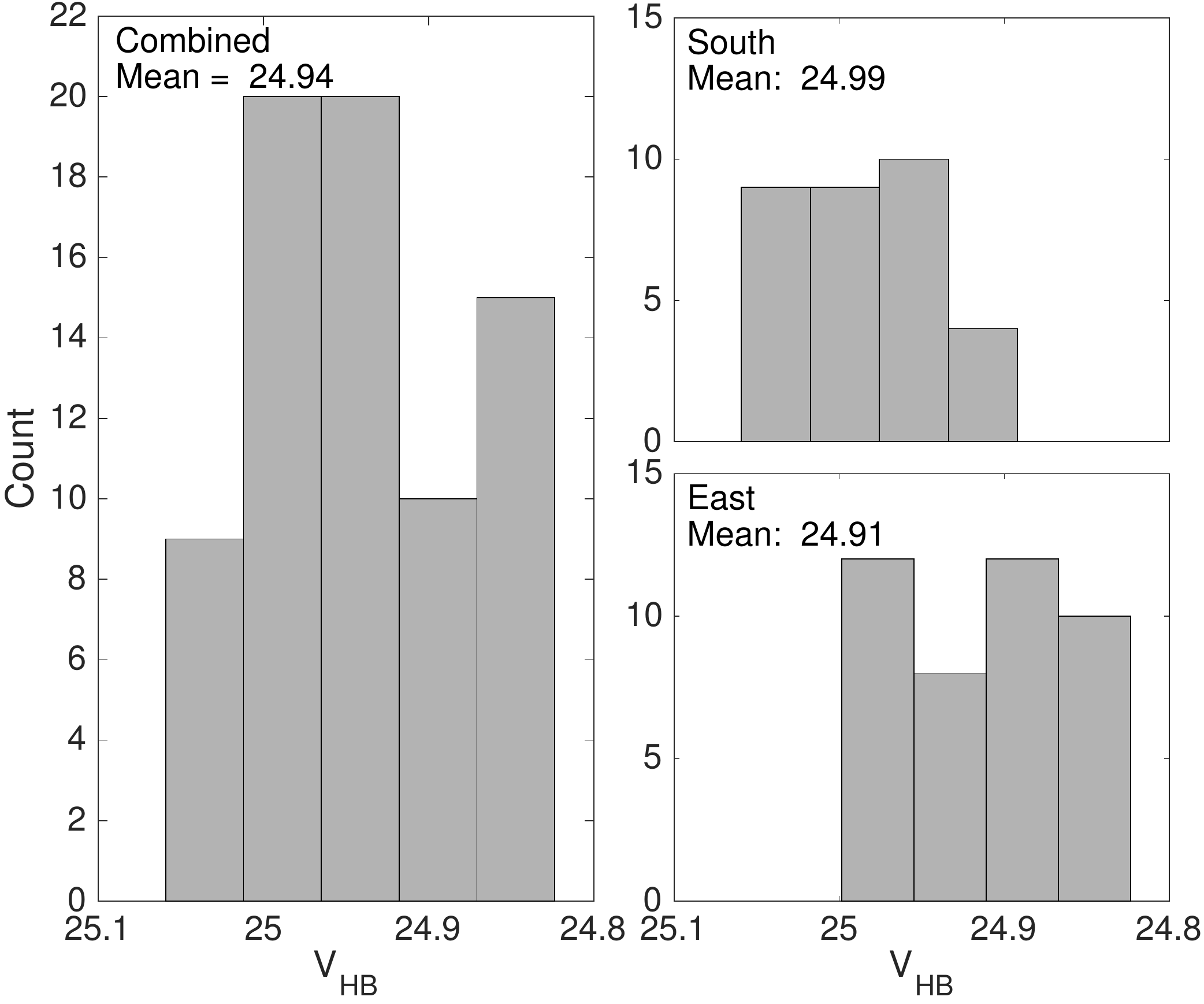}
  \caption[Histogram of calculated $V_{\mathrm{HB}}$ values]{Distribution of calculated $V_{\mathrm{HB}}$ values for the East field stars, South field stars, and the combined distribution. Panels are marked accordingly, and the mean of each range is displayed; errors in the mean are \les $0.02$.}
  \label{fig:VhbHist}
\end{figure}

The line of sight reddening values thus derived range from $\rm E(B-V)=0.19$ -- $0.33$, with mean values of $\rm E(B-V)=0.24$ and $\rm E(B-V)=0.28$ in the East and South fields, respectively. This serves to re-emphasis the spatially variable nature of the dust reddening within and in front of NGC~6822. The average synthesized HB magnitude across both the East and South field was $\langle V_{\mathrm{HB}}\rangle= 24.94\pm0.03$ with standard deviation $\sigma=0.06$. The resultant distribution of $V_{\mathrm{HB}}$ can be seen in Figure \ref{fig:VhbHist}. As expected, the East field is less highly reddened than the South field, where the HI column density and IRAS emission are both higher. If the two fields were analysed assuming a single reddening value, then stars of equal apparent magnitude in the two fields would not correspond to stars of equal absolute magnitude, and the influence of surface gravity on CaT equivalent width would yield a small systematic error in the metallicity measurements.

\begin{figure}
  \centering
\includegraphics[width=0.45\textwidth]{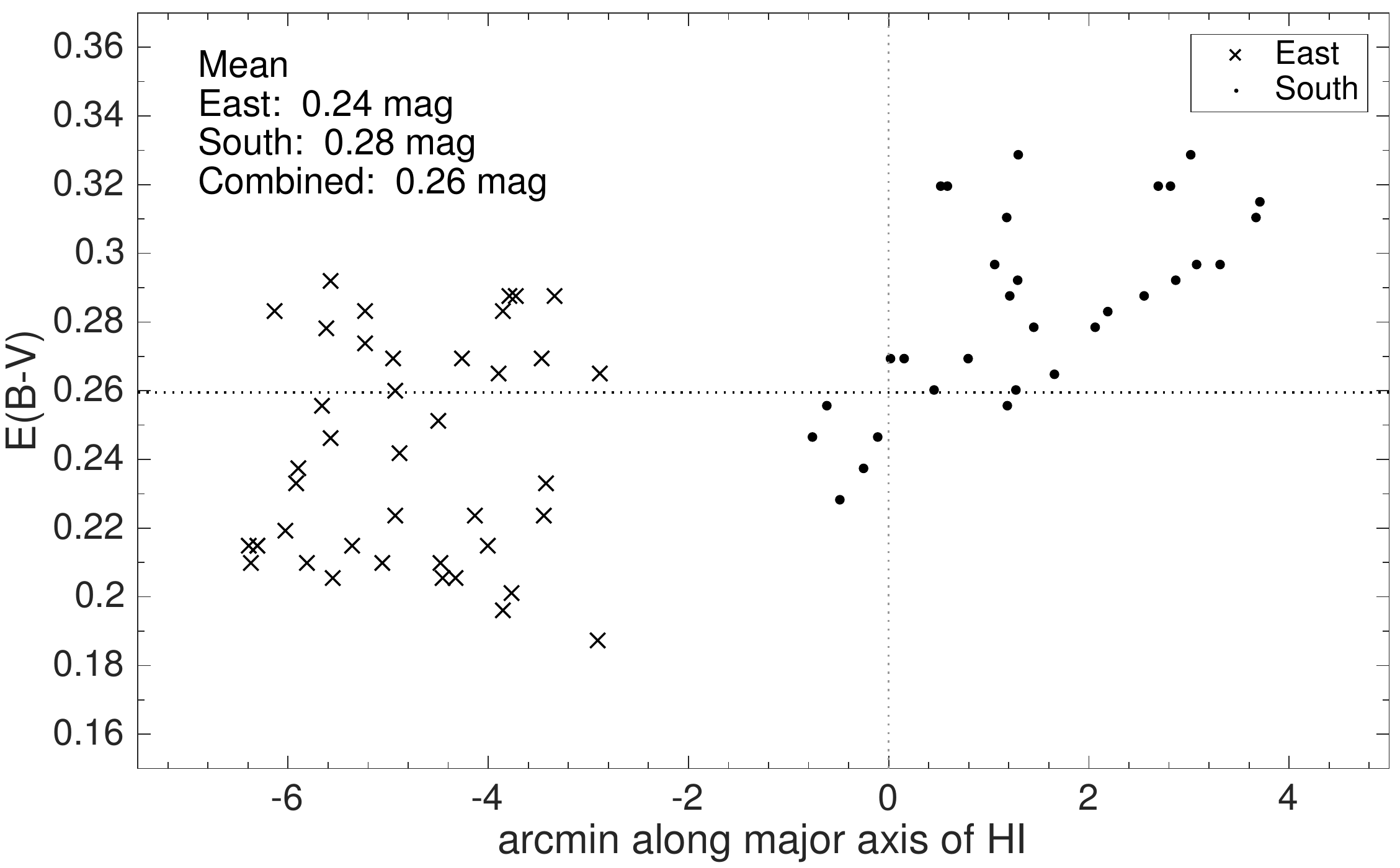}
  \caption[]{Reddening variation across the 130\degrr{} PA of the \HI{} envelope. Individual errors in the calculated extinction are $\leq 0.04$.  The dashed lines show the radial zero point and the average reddening for the combined data for convenience. The mean extinction value for each field is displayed, the error in the mean is 0.02~mag.}
  \label{fig:redd_distr}
\end{figure}

Since we have used the same equivalent width measurement technique as \cite{Cole2004}, and the same linearisation of the equivalent width, we will need to use the appropriately determined linear relation between the reduced equivalent width and metallicity index, [Fe/H]. 
\begin{align} 
\mathrm{[Fe/H]}&=(0.362\pm0.014)W'-(2.966\pm0.032)  \label{eq:FeH}
\end{align}
This empirical fit was derived by \cite{Cole2004} over a metallicity range  $-2.0\lesssim\mathrm{[Fe/H]}\lesssim -0.2$ and is estimated to representative of actual [Fe/H] conservatively to within 15\% over this range for RGB stars. { Between the rms
scatter around the best-fit line from \citet{Cole2004} and the uncertainties in the high-dispersion metallicity scale for calibrating clusters, we conservatively estimate a possible systematic error of up to 0.25~dex.}

\begin{figure}
  \centering
\includegraphics[width=0.45\textwidth]{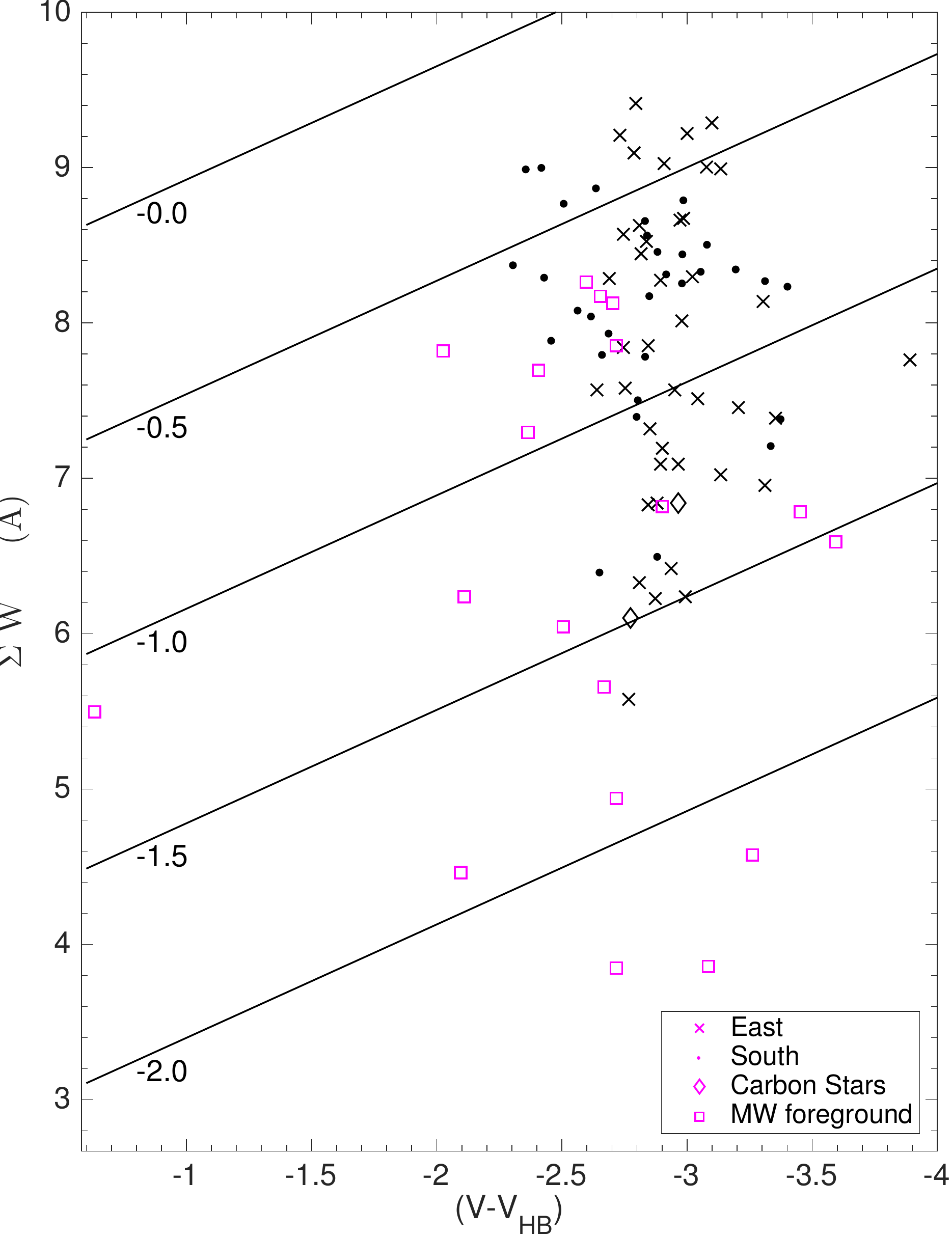}
  \caption[Summed equivalent widths of the calcium triplet ($\Sigma W$) vs. $V$ magnitude distance from $V_{\mathrm{HB}}$]{Distribution of summed equivalent widths of the calcium triplet ($\Sigma W$) as a function of $V$ magnitude distance from $V_{\mathrm{HB}}$. The solid lines represent constant [Fe/H] metallicity according to the calibration by (\citealt{Cole2004}; see equation \ref{eq:FeH}). }
  \label{fig:SumW_delV}
\end{figure}

\begin{table}
  \centering
  \caption[Summary of systematic and random errors]{Summary of systematic and random errors for the most important variables in this paper.}
    \begin{tabular}{rccc}
    \hline
    \hline
    \multicolumn{1}{c}{\textbf{[Fe/H]}} & \textbf{Random} & \textbf{Systematic} & \textbf{Adopted} \\
    \hline
    \multicolumn{1}{l}{$V$} & 0.03  & - $^{\dagger}$ & 0.03 \\
    \multicolumn{1}{l}{$V_{\mathrm{HB}}$} & 0.07  & 0.2   & 0.21 \\
    \multicolumn{1}{l}{$v_{rad}$ (km s\per{})} & 1.2   & 6.5   & 6.61 \\
    \multicolumn{1}{l}{[Fe/H]} & 0.03  & 0.25  & 0.25 \\
    \multicolumn{1}{l}{$\Sigma W$ (\AA)} & 0.08  & 0.09  & 0.12 \\
    \hline
    \end{tabular}%
    \hspace{14px} $^{\dagger}$ {\small $V$ magnitudes for our stars were converted from \cite{McMahon2001} photometry, whilst random errors are quoted we have no grounds on which to estimate a systematic error.}
  \label{tab:errors}%
\end{table}%
% section calculation_of_ (end)
 
\section{Milky Way foreground contamination} % (fold)
\label{sec:milky_way_foreground_contamination}
Due to our target selection by broadband photometry and the location of NGC~6822, our spectroscopic sample suffered from heavy contamination by foreground MW stars. Since our targets are within a magnitude range 21 \les{} $V$ \les{} 23 and are selected to have colours characteristic of K stars, then any foreground contaminants will be K dwarfs. To filter these contaminating dwarfs from our sample of red giants, we can exploit  the surface gravity dependence of key spectral lines
of sodium and magnesium together with the calcium triplet.

\begin{figure}
    \centering
    \includegraphics[width=0.45\textwidth]{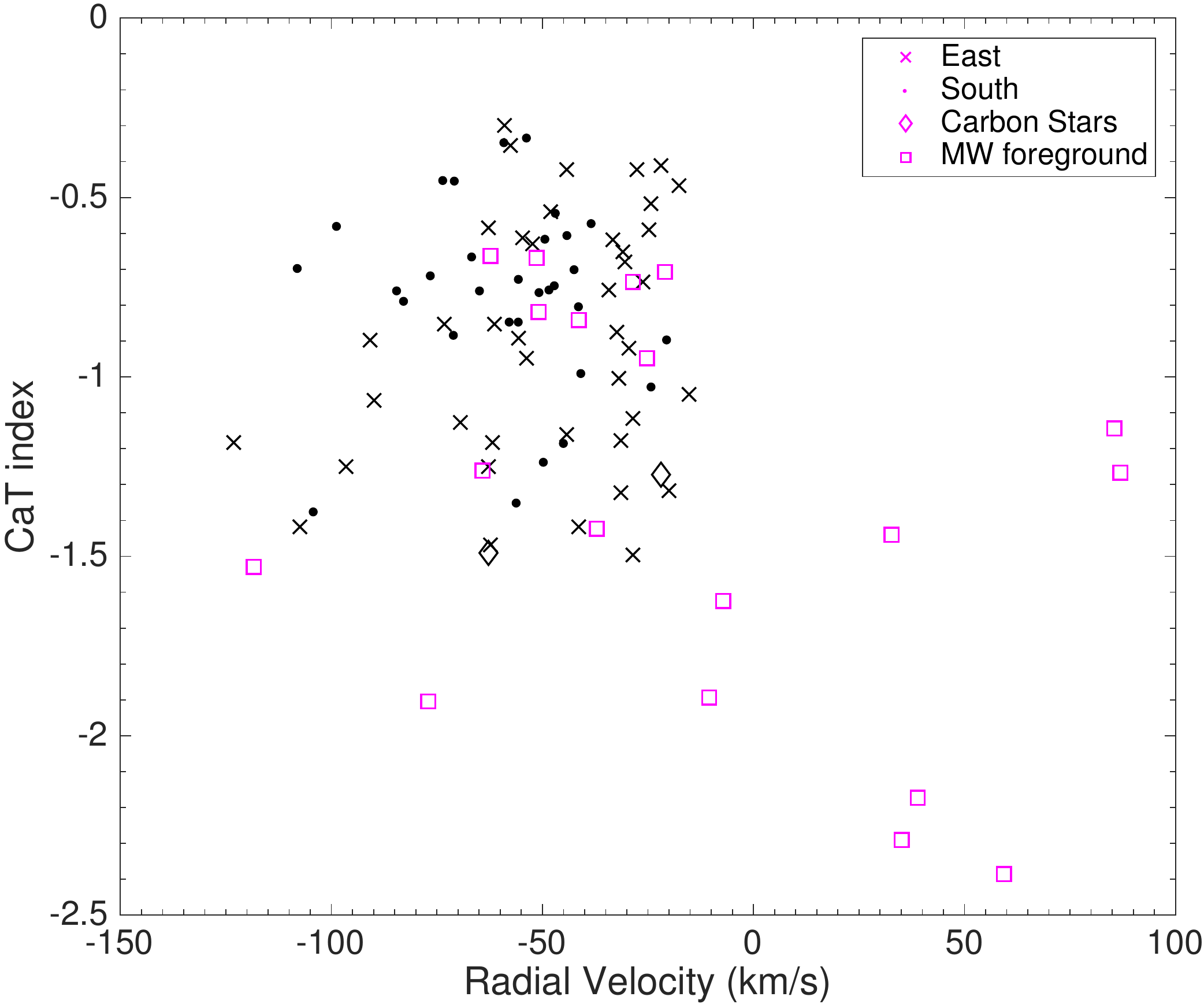}
    \caption[Distribution of CaT index vs. heliocentric velocity for our target stars]{Distribution of CaT index vs. heliocentric velocity for our target stars. The stars are segregated into South field stars, East field stars, Carbon stars, and MW foreground stars. Note that the CaT index of the RGB stars (marked `East' and `South') can be viewed as a metallicity (for K giants only) if desired. The stars marked as MW foreground are the final results of our decontamination.}
    \label{fig:metVrad}
\end{figure}

Once equivalent widths and velocities were acquired for our stars, we constructed a calcium \textsc{ii} triplet index (CaT index) using the metallicity calibration outlined in \S{} \ref{sec:calculation_of_met} (see also \citealt{Cole2004}). Whilst this translates directly to an [Fe/H] metallicity for our RGB targets, the calcium triplet calibration for metallicity does not extend to dwarf stars. Figure \ref{fig:metVrad} is a plot of the distribution of CaT indices versus radial velocity. Since we expect that MW foreground stars will be shifted towards less negative velocities, stars with $v_{rad}$ $>$ 0 km s\per{} are likely to be contaminants. Similarly, those stars with low CaT indices are more likely to be contaminants. Acknowledging this, we conservatively marked stars with $v_{rad}\,\gtrsim$ 20 km s\per{} or CaT index more than three standard deviations from the mean as potential contaminants for reference in later filtration methods.

\cite{Diaz1989} showed the dependence of the Mg \textsc{i} line at $\lambda$ = 8807 \AA{} on surface gravity and temperature. \cite{Zhou1991} similarly investigated the variation in the Na \textsc{i} $\lambda$ = 8183 \AA{}, 8195 \AA{} doublet with star type. Applying this, we can expect for dwarf stars contaminating our RGB sample that the Mg \textsc{i}, and Na \textsc{i} doublet line strengths will be much greater than in our NGC~6822 RGB members for a given CaT equivalent width \citep{Battaglia2012}. Whilst these line strengths will still vary with metallicity, in our sample we can expect that the difference in surface gravity between our star classes will dominate. Disappointingly, the Na \textsc{i} doublet sits outside of the wavelength range for our targets closest to the red edge of the CCD. Each of the 91 non-carbon star spectra were therefore merely checked by eye and only the stars with the strongest Na \textsc{i} doublet lines marked as potential contaminants.

Since the Mg \textsc{i} line was present in all of our spectra, we measured the equivalent width in the manner outlined in \S{} \ref{sec:equivalent_width_measurements} using bandpasses defined in \cite{Diaz1989}. Additionally, because the template stars used for the Fourier cross-correlation radial velocity measurements were of known stellar type (23 RGB stars, and 1 dwarf star)  these made excellent references for decontamination. As no equivalent width measurements had previously been made on these templates, the Ca \textsc{ii} $\lambda$ = 8542 \AA{}, 8662 \AA{} equivalent widths were measured along with the Mg \textsc{i} $\lambda$ = 8807 \AA{} line (see Table  \ref{tab:templates}) for the purpose of this filtration.

\begin{figure}
    \centering
    \includegraphics[width=0.45\textwidth]{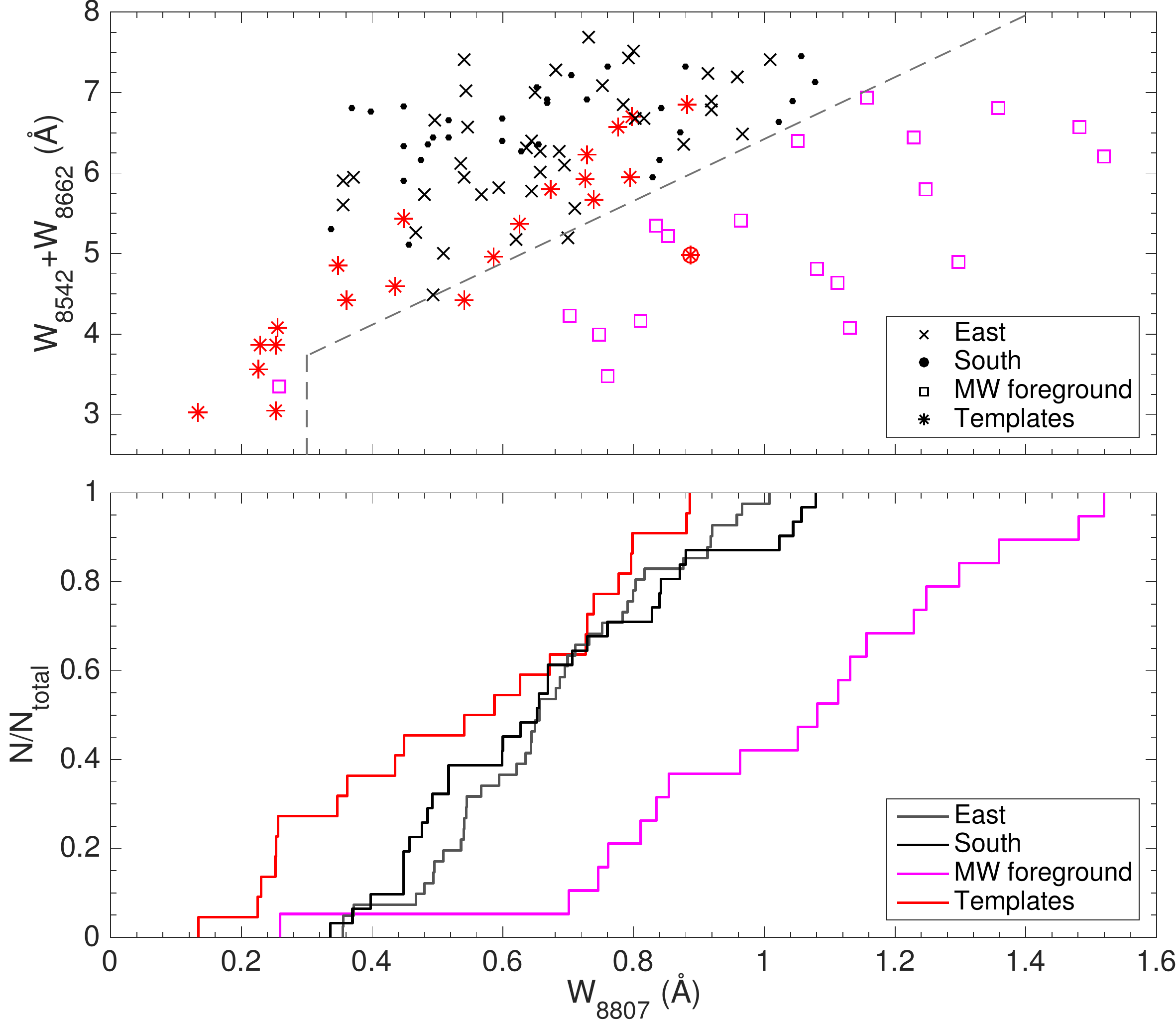}
    \caption[Test for foreground contaminants: W$_{8542}$ vs. W$_{8807}$]{Top panel: Distribution of the equivalent width of the $\lambda=8807$ \AA{} Mg \textsc{i} line versus the sum of the equivalent widths of the Ca \textsc{ii} $\lambda= 8542$ \AA{} and 8662 \AA{} lines. The points are subcategorized into East field stars, South field stars, MW foreground stars, and template RGB stars; these are labeled accordingly. The template star marked with a red circle is the dwarf star from the template sample. The MW foreground star with the smallest W$_{8807}$ was rejected based on radial velocity ($v_{rad} = $+$35.1$ km s\per{}). This is the primary plot that was used for filtering the foreground contaminants. { The dashed line is the empirical separation of dwarf stars and red giants suggested by \cite{Battaglia2012}.} Bottom panel: cumulative distribution function of the top panel; note the similarities in gradients and the shift of the foreground contaminants to higher Mg \textsc{i} line widths. Errors are no more than $\pm$ 0.1 \AA{} on either axes.}
    \label{fig:mgTest1}
\end{figure}

Figure \ref{fig:mgTest1} is a plot of the distribution of Mg \textsc{i} line strengths versus the sum of the line strengths of the $\lambda$ = 8542 \AA{} and 8662 \AA{} Ca \textsc{ii} lines. Even without any radial velocity information, two distinct populations in the plot are clearly visible. The presence of the dwarf template star in the population with broader Mg \textsc{i} lines suggests that the other stars with strong Mg \textsc{i} 8807 lines are foreground contaminants. This is confirmed by cross-correlating the 8807-strong stars with the previously identified radial velocity outliers and the Na \textsc{i} doublet observations where available. Culling by Mg \textsc{i} gives a stricter foreground rejection than by radial velocity alone owing to the strong velocity overlap of NGC~6822 with the Milky Way.

{ As a further confidence check, the \cite{Battaglia2012} empirical dwarf and red giant star separation line is over-plotted onto Figure \ref{fig:mgTest1}. This matches well with the eliminations we performed using first pass grouping with the known template stars, and radial velocity checks. The addition of the line does bring to light two exceptions to the Battaglia \& Starkenburg division. Of particular note is the template star that is a known red giant but appears slightly on the dwarf side of the dividing line. This confirms the uncertainty of the empirical division close to the line, and particularly at lower metallicity. Because of this we have left all of our borderline classifications unaltered. The other exception is a star marked as a foreground contaminant on the giant side of the dividing line. This is most likely a dwarf star due to its radial velocity ($v_{rad} = +35.1$ km s\per{} -- \apr{} 90 km s\per{} greater than the bulk motion of NGC~6822) and low metallicity ([Fe/H]~=~$-2.29$), this star is therefore left out of our red giant sample.}

{ The \cite{Battaglia2012} division would have sufficiently eliminated our dwarf star contamination, but based on our small sample it might be slightly too aggressive. Having added the line after our final decisions on membership had been made, we are confident that our elimination of Milky Way dwarfs was thorough. Further discrimination between the two { star types} cannot be achieved without the use of other filtration methods, such as large sample photometric analysis (\citealt{Gullieuszik2008}; \citealt{Sibbons2011}). Looking at the scatter in equivalent widths in Figure \ref{fig:mgTest1} the S/N limitation of our data becomes apparent. If our data were of higher S/N, the distinction between RGB and dwarf contaminants might be more apparent and the leakage across the Battaglia \& Starkenburg line possibly reduced.} 

That being said, this method has proven to be a simple yet adequate method for removing foreground contamination of dwarf stars from a relatively low S/N spectral sample of K type red giants. If more of our sample extended to the wavelength of the Na \textsc{i} doublet, the identification of foreground contaminants would be even stronger, but we have shown that the sodium lines are not a requirement for rejection by surface gravity. This technique could also be extended to include comparisons with stellar atmosphere models such as those in \cite{Kirby2012} in situations where doubt remains as to the efficiency of dwarf/giant separation.

% \begin{figure}
%     \centering
%     \includegraphics[width=0.45\textwidth]{MgTest2-eps-converted-to.pdf}
%     \caption[Test for foreground contaminants: W$_{8542}$/W$_{8807}$ vs. W$_{8807}$]{Top panel: { Distribution of the sum of the equivalent widths of the Ca \textsc{ii} $\lambda= 8542$ \AA{} and 8662 \AA{} lines scaled by the width of Mg \textsc{i} versus the equivalent width of Mg \textsc{i}.} isThe symbols are as in Figure~\ref{fig:mgTest1}. This is the  plot that was used for a confidence check of the previous contamination filter. { The dashed line is the empirical separation of dwarf stars and red giants suggested by \cite{Battaglia2012} altered appropriately for the new coordinates.} Bottom panel: cumulative distribution function of the top panel. Note the clustering of the contaminants towards higher Mg \textsc{i} line strengths (low W$_{8542}$/W$_{8807}$). Errors are no more than $\pm$ 0.1 \AA on the vertical axis, and $\pm$0.2 on the horizontal axis.}
%     \label{fig:mgTest2}
% \end{figure}

{ Contrastingly, \cite{Tolstoy2001} used only the stellar radial velocity to reject foreground stars; \citet{Kirby2013} and \citet{Kirby2014} also searched their member list for contaminants using the Na~\textsc{I} doublet but did not find any. If we had followed this procedure for the present sample, we would have only identified 6 of our 18 contaminants.}
% section milky_way_foreground_contamination (end)

\section{Stellar dynamics of NGC~6822} % (fold)
\label{sec:stellar_dynamics_of_ngc_6822}

\begin{figure}
  \centering
 \includegraphics[width=.45\textwidth]{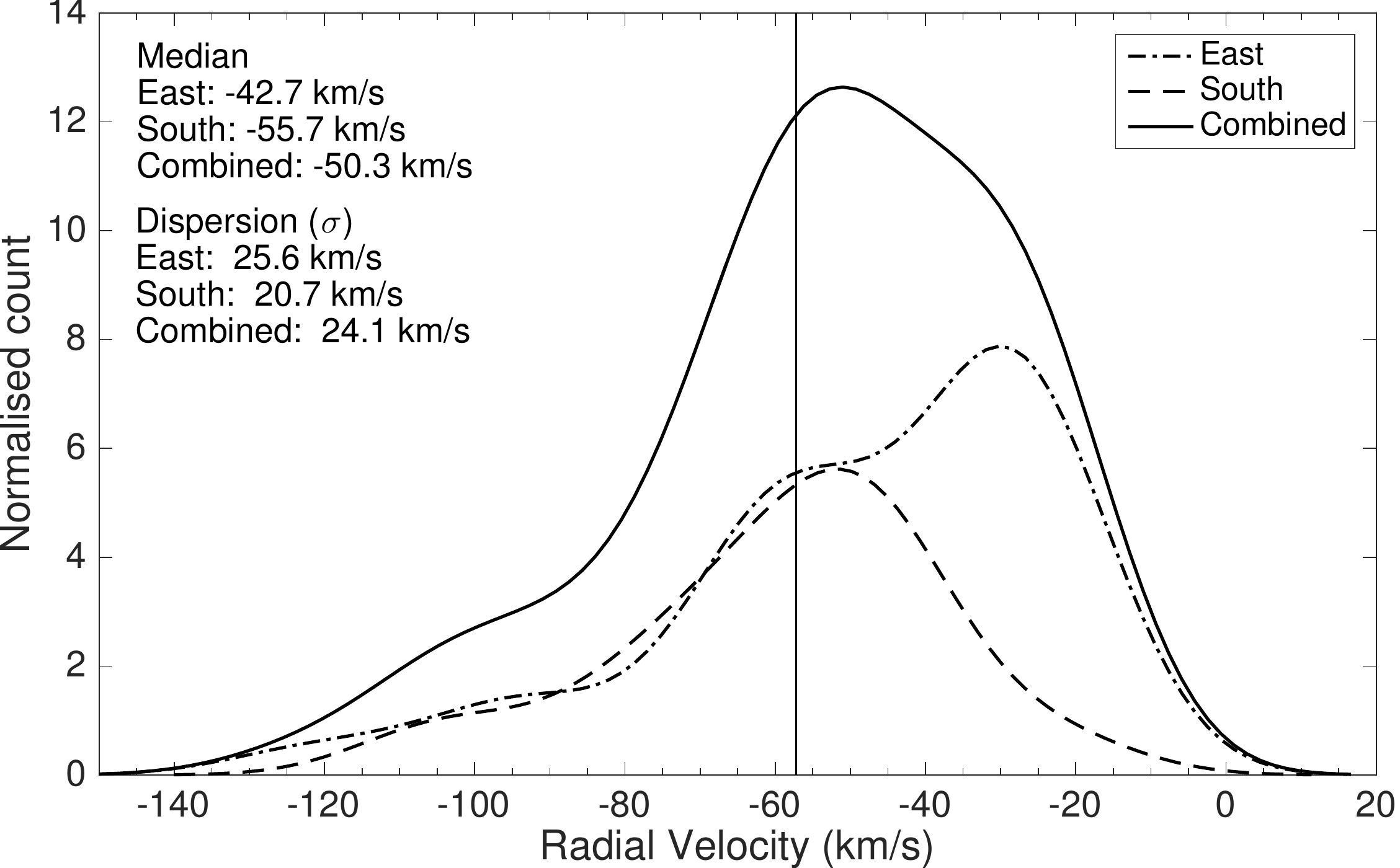}
  \caption[Velocity distribution of NGC~6822 RGB members]{Distribution of velocities of the 74 NGC~6822 member RGB and carbon stars. Shown is a probability distribution function (PDF) of velocities normalized to the total count for each PDF. The combined distribution and the separate distributions of East and South targets are both shown. The total number of members in the East field is now 42, with the remaining 32 belonging to the South field - this includes the two previously identified carbon stars. The median velocity for each PDF is displayed without uncertainty, the velocity dispersions are also shown. The solid vertical line represents the average \HI{} velocity of NGC~6822 according to \cite{Koribalski2004}: $\sim -57.2$ km s\per{}. }
  \label{fig:cleanVdist}
\end{figure}

The radial velocity distribution for our 72 confirmed member RGB stars and 2 Carbon stars is shown in Figure \ref{fig:cleanVdist}. The combined data for both fields give $\langle v_{rad}\rangle=-53.0 \pm 2.2$ km s\per{}, and $\sigma_{v}=24.0$ km s\per{}. 
{ The median velocity of the east field is 13~km~s\per{} more positive than the south field, which as discussed below could be taken as evidence for rotation, albeit with a high degree of dynamical support by velocity dispersion. The velocity distributions also appear different; Figure \ref{fig:cleanVdist} shows a hint of bimodality, stronger in the East field than in the South. The shift in the median of the East field could be due to the stronger contribution of stars in a population which, if present in the South field, is much weaker.}
A Kolmogorov-Smirnov (KS) test \citep{Stephens1979} p-value of 0.0022 confirms that the East and South field velocities can be treated as independent distributions. 

The first study for which velocities of individual intermediate-age and old stars were determined was conducted by \cite{Tolstoy2001}. This study used much of the same data acquisition and spectral analysis techniques as the present paper, but with only a limited number of stars. They acquired a mean velocity from 21 member stars of $\langle v_{rad}\rangle=-60.1$ km s\per{} with a dispersion $\sigma_{v}=24.5$ km s\per{}. \cite{Kirby2013} in a spectroscopic study of 292 targets found $\langle v_{rad}\rangle=-54.5 \pm 1.7$ km s\per{}, and $\sigma_{v}=23.2 \pm 1.2$ km s\per{}. They found the velocity dispersion to be by far the highest of the Local Group dwarf irregular galaxies in their sample, although this is still not as high as the radial velocity dispersion of the similarly luminous SMC.  The relatively high line of sight velocity dispersion and { comparatively small velocity difference that could be attributable to rotation} suggests the possibility that the RGB population of NGC~6822 is more well-described as a dynamically hot, spheroidal distribution than as a rotationally supported disk galaxy. 

If we assume that the \HI{} and stellar content are dynamically similar over the inner 10$'$, then the \citealt{Weldrake2003} velocity gradient can be used to approximate the rotational velocity of the stellar population in this region. Accepting this, we get $v_{rot}/\sigma_{v}$ = 24/24.1 \apr{} 1 for $\abs{r_{ell}} \lesssim$ 1.4 kpc. As a comparison, \cite{Leaman2012} found an identical $v_{rot}/\sigma_{v}$ for WLM over the inner 6$\,'$ (1.6 kpc; see Figure 11 in their paper). Comparing this directly to the same quantities but instead using the \HI{} velocity dispersion from \cite{Weldrake2003} of \apr{} 8 km s\per{} we get  $v_{rot}/\sigma_{v}$ $\simeq$ 4 for the inner 10$\,'$, and  $v_{rot}/\sigma_{v}$ $\simeq$ 7 as an upper limit. Since the stars were originally formed from the \HI{}, this provides clear evidence of a decoupling between the intermediate-age stars and the current gas distribution: either dynamical heating has taken place over the stellar lifetimes, or new gas with different energy and angular momentum has been accreted into NGC~6822 in the intervening time.

Continuing with the use of the \citealt{Weldrake2003} \HI{} velocity gradient, we can predict the approximate velocity that objects at the location of these fields should have, given their projected separation along the 130\degrr position angle of the gradient. For the South and East field we expect velocities of $-51\pm1.6$ km s\per{} and $-41\pm1.6$ km s\per{} respectively. These predicted velocities match well with the $-50$ km s\per{} peak in the over-all velocity distribution of our sample as well as the median values for each of the subfields. The East field however presents an interesting example; whilst there is a significant peak in this distribution at $\sim -50$ km s\per{} most of the objects had velocities $\sim -30$ km s\per{}. It is impossible to discern the nature of this anomaly without further data; it could be merely due to chance sampling and our small sample size, or a manifestation of decoupled sub-population with a different systemic velocity. 

%section stellar_dynamics_of_ngc_6822 (end)

\section{Metallicity analysis} % (fold)
\label{sec:metallicity_analysis}

\begin{figure}
    \centering
    \includegraphics[width=0.45\textwidth]{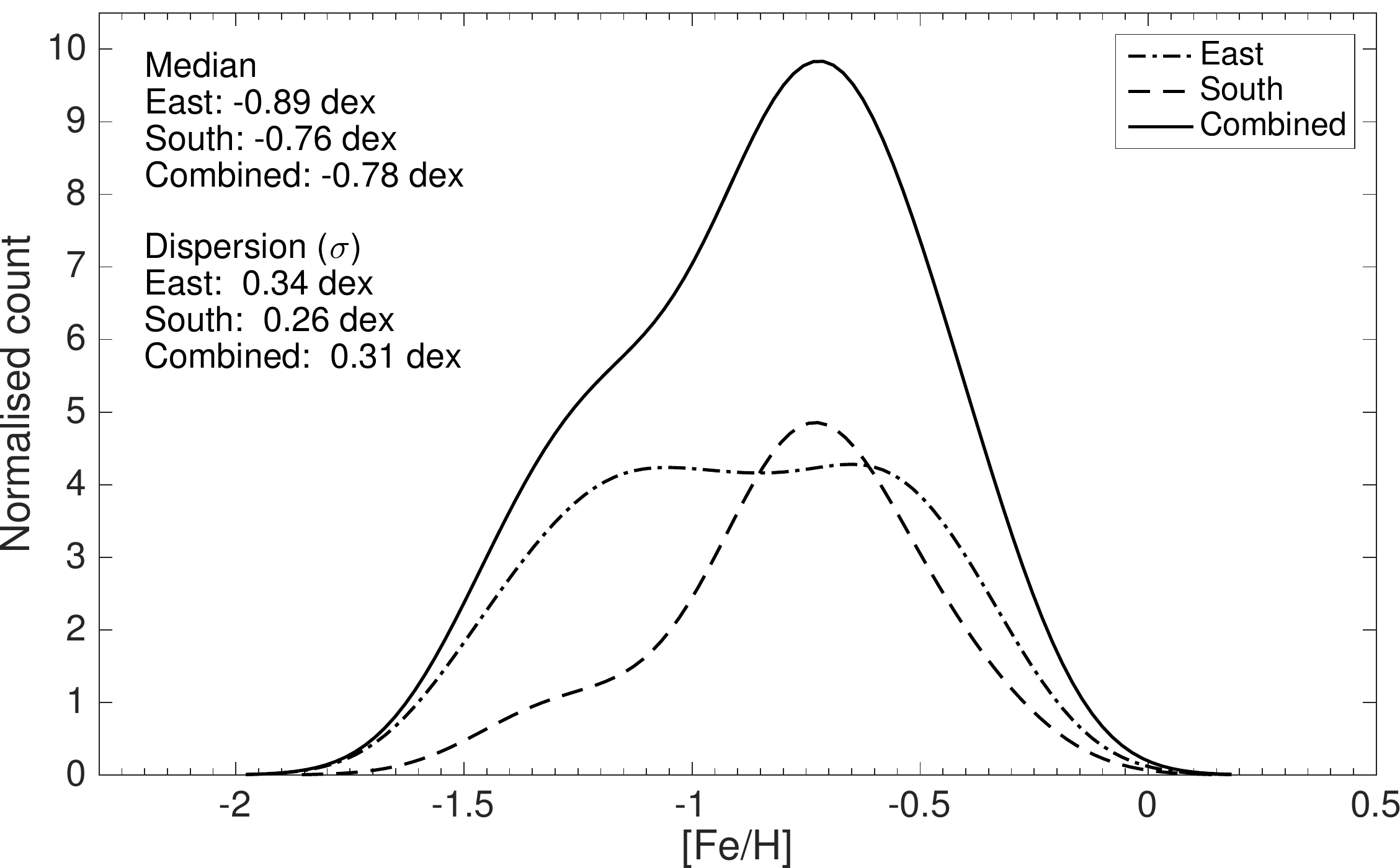}
	\caption[Metallicity distribution function of our 72 red giants]{Metallicity distribution function (MDF) of the 72 red giant stars in our sample. Shown is a probability distribution function (PDF) of velocities normalized to the total count for each PDF. These have been separated into East and South targets, as well as the combined distribution being shown. The total number of members in the East field is now 41, with the remaining 32 belonging to the South field - this does not include the two previously identified carbon stars. The median metallicities and dispersions are also shown.}
    \label{fig:met_KDE}
\end{figure}

Taking the mean metallicity of our data we find \mean{[Fe/H]} $=-0.84\pm0.04$ with a dispersion of $\sigma=0.31$ dex. Figure \ref{fig:met_KDE} shows metallicity distribution function (MDF) of our 72 red giant stars. A cursory inspection reveals the presence of a significant low-metallicity tail in the East field data that is suppressed or not present in the South field. While the South field has a similar shape to other dwarf irregular galaxies, the East field MDF is more flat-topped; the means differ significantly, by $\approx$0.13~dex. With such a small sample of red giants, a KS p-value of 0.057 is arguably inconclusive, but does suggest that we have to be careful when considering the metallicities of the South and East field as being drawn from independent distributions.

In table \ref{tab:metals} we have summarized the published [Fe/H] metallicities of NGC~6822, including measurement technique, citation, and the approximate age of the target population; the table is ordered by age of tracer. 

\begin{table*}
  \centering
  \caption[Summary of literary metallicities for NGC~6822]{\centering Summary of literary [Fe/H] for NGC 6822, approximate ages of targets are shown.}
    \begin{tabular}{rccr}
    \hline
    \hline
    \multicolumn{1}{c}{\textbf{[Fe/H]}} & \textbf{Tracer} & \textbf{Approx. age} & \multicolumn{1}{c}{\textbf{Reference}} \\
    \hline
    \multicolumn{1}{l}{$-0.6\pm0.2$} & \HII{} regions & \les{} 10 Myr & \cite{Skillman1989} \\
    \multicolumn{1}{l}{$-0.5\pm0.2$} & B-Supergiants spect. & \les{} 300 Myr & \cite{Muschielok1999} \\
    \multicolumn{1}{l}{$-0.49\pm0.22$} & A-Supergiants spect. & \les{} 500 Myr & \cite{Venn2000} \\
    \multicolumn{1}{l}{$-0.79\pm0.15$} & IL & $0.01-1$ Gyr & Colucci \&{} Bernstein (\citeyear{Colucci2011}) \\
    \multicolumn{1}{l}{$-1.25\pm0.5$$^{\dagger}$} & AGB C/M ratio & $0.5-5$ Gyr     & \cite{Kacharov2012} \\
    \multicolumn{1}{l}{$-1.14\pm0.08$} & AGB C/M ratio & $0.5-5$ Gyr & \cite{Sibbons2011} \\
    \multicolumn{1}{l}{$\sim{}-1.0$} & IL spectra of GC &  $\sim{}2$ Gyr & \cite{Cohen1998} \\
    \multicolumn{1}{l}{$-1.0\pm0.3$} & Ca \textsc{ii} triplet (RGB) & $2-5$ Gyr & \cite{Tolstoy2001} \\
    \multicolumn{1}{l}{$-0.84\pm0.04$} & Ca \textsc{ii} triplet (RGB) & $2-5$ Gyr & This work \\
    \multicolumn{1}{l}{$-0.98\pm0.01$$^{*}$} & Spectral synthesis & $2-5$ Gyr & Evan Kirby Priv. Comm. (2015) \\
    \multicolumn{1}{l}{$-1.05\pm0.01$} & Spectral synthesis & $2-5$ Gyr & \cite{Kirby2013} \\
    \multicolumn{1}{l}{$-1.5\pm0.03$} & RGB slope & 6$\pm$3 Gyr     & \cite{Gallart1996a} \\
    \multicolumn{1}{l}{$-1.61\pm0.02$} & IL spectra of GC &  \gre{} 7 Gyr & Colucci \&{} Bernstein (\citeyear{Colucci2011}) \\
    \multicolumn{1}{l}{$-1.95\pm0.15$} & IL spectra of GC &  $8-15$ Gyr & \cite{Cohen1998} \\
    \hline
    \end{tabular}%
    \hspace{194px} $^{\dagger}$  {\small spectra of globular cluster}\\
    \hspace{4px} $^{*}$ {\small Recalculated by Evan Kirby matching the assumed reddening to that of our East field}
  \label{tab:metals}%
\end{table*}%

The most direct comparisons for our data are the RGB abundances obtained by \cite{Tolstoy2001}, and \cite{Kirby2013}. \cite{Tolstoy2001} used a virtually identical method to us, but obtained lower signal to noise and a smaller sample size. Thus it is not surprising that our mean abundances agree within the errors. \citet{Kirby2013} obtained a much larger sample using the Keck/DEIMOS spectrograph, analysing a large number of stars along the central bar of NGC~6822 and calculating [Fe/H] by spectral synthesis matched to Fe~I lines. Their sample has a high degree of spatial overlap with our South field and with the more northerly field analysed by \citet{Tolstoy2001}.  Their sample extends to bluer colours and fainter magnitudes than does our own.  

Our East field covering the region of the HI hole was not observed by either \cite{Tolstoy2001} or \cite{Kirby2013}, and contains proportionately more of the low-metallicity stars.  However, our derived average metallicity is slightly higher than either previous study. The source of the discrepancy may be traceable to the stricter foreground rejection employed here, to the quite different metallicity derivations from the spectra (comparing to Kirby), or partially to the individual line of sight reddening corrections (comparing to Tolstoy). In general, the red giant results tend to support the conclusion of \cite{Muschielok1999} and
others that NGC~6822 has approximately the same or slightly higher metallicity as the SMC.  

\citet{Tolstoy2001} assumed a constant reddening of E$(B-V)=0.24$ based on the \cite{Schlegel1998} dust maps, while \citet{Kirby2013} made the nearly equivalent assumption that E$(B-V)=0.25$ based on the results from \cite{Massey2007}. In contrast, a more recent survey by \cite{Fusco2012} found E$(B-V)=0.30\pm0.02$ and 0.37$\pm0.03$ for an exterior and central field, respectively, slightly higher than our derived average values.  We can investigate the effects of the different reddening assumptions on previous RGB metallicity estimates in NGC~6822.

\begin{figure}
    \centering
   \includegraphics[width=0.49\textwidth]{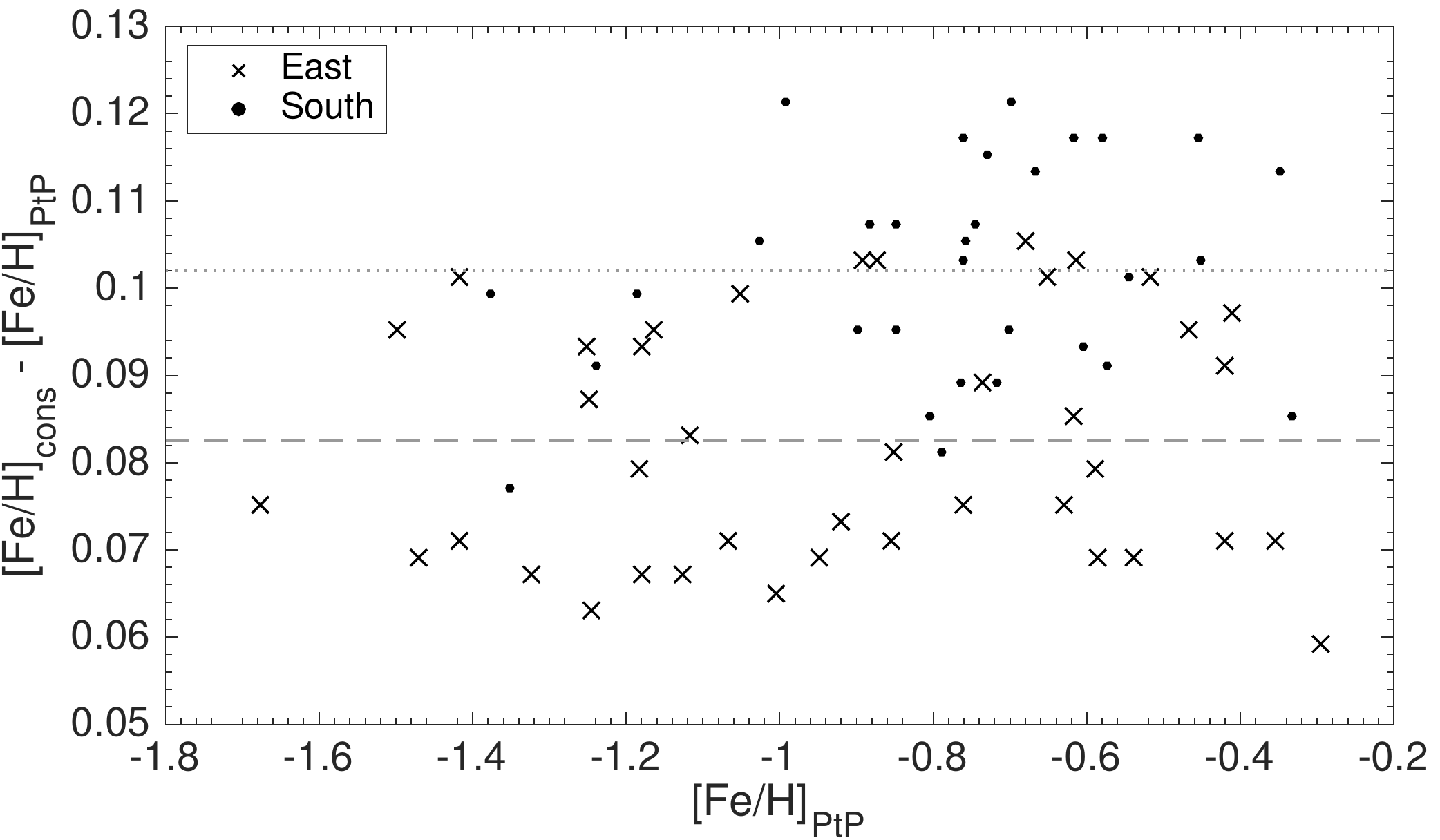}
    \caption[]{Comparison of the estimated metallicities using the regional (point-to-point) dereddening ([Fe/H]$_{\rm PtP}$) in this work and the constant dereddening E(B$-$V) = 0.24 employed by Tolstoy et al. (\citeyear{Tolstoy2001}; [Fe/H]$_{\rm cons}$). Individual errors in metallicity are 0.3 dex; mean differences between the techniques are plotted separately for each field.}
    \label{fig:metal_comparison}
\end{figure}

\citet{Tolstoy2001} used a very similar methodology to this paper, but assumed a constant value of $V_{\rm HB}$ = 24.6 based on HST/WFPC2 data \citep{Wyder2001}. Figure \ref{fig:metal_comparison} shows the shift in our metallicities if we had assumed the same constant reddening and HB magnitude as \cite{Tolstoy2001}. The mean shift in metallicity across both fields is $-0.09$ dex, very nearly eliminating any discrepancy with the results of the earlier paper. The plot also shows that, as expected, there is no metallicity-dependent offset introduced by our point-to-point reddening correction. The { significance of the detected metallicity difference between the South and East fields would have been significantly decreased by the reduction in apparent mean difference and the increase in apparent dispersion} between fields if a constant E(B$-$V) had been used.

It is more difficult to make a similar quantitative comparison to the \citet{Kirby2013} results without re-interpolating within their grid of synthetic spectra, but we can attempt to infer the likely effects of assumed reddening on the model atmosphere parameters T$_{\mathrm{eff}}$, log$g$, and [Fe/H] from their spectral fitting. By subtracting a higher reddening and extinction from the photometry, { we predict that Kirby et al.\ would have found a higher effective temperature and a very slightly lower surface gravity, bringing the \citet{Kirby2013} results into closer agreement with our own. Fortunately, Evan Kirby, the referee of this paper, was kind enough to recalculate metallicity from his data assuming a reddening of E(B$-$V) = 0.30, consistent with the value in our eastern field. He reports an increase in metallicity of 0.07~dex for a reddening increase of 0.05 mag. We conclude that the different reddening estimates contribute to the different results of the two studies, but are comparable to or less than the differences resulting from the different measurement techniques.}

Since neither \cite{Tolstoy2001} nor \cite{Kirby2013} performed a non-uniform reddening correction, the brightness of their most reddened stars would have been underestimated resulting in an underestimation of the metallicities of these stars. This is exacerbated by the tendency for the inner-most stars of a galaxy to be both more highly reddened and more metal rich. As a result, our study is more sensitive to the presence of metallicity gradients in NGC~6822 than either previous study. This sensitivity is likely to be enhanced if mixing along the bar has homogenized metallicities in the central regions, because our East field lies further from the bar than either previous study.

{ Weighing against this is the possibility that the metallicity dependence of the TRGB magnitude has introduced additional complexity that has not adequately been accounted for. Because our dereddening technique uses relative magnitudes, we are not subject to errors in the absolute calibration of the TRGB magnitude. However, the intrinsic metallicity dependence of I$_{\mathrm{TRGB}}$ could change the results of our analysis given that the south and east fields differ in mean [Fe/H] by $\approx$~0.13~dex. According to the calibration in \cite{Bella2001}, this difference is expected to produce a difference in TRGB magnitude of 0.032 (with the eastern field brighter); if this difference was misinterpreted as a contribution to A$_V$ in the south field, the additional spurious reddening would amount to E(B$-$V) $\approx$0.01~mag. This is clearly well below the threshold of significance for our analysis.}

% section metallicity_analysis (end)

\section{Discussion}

\subsection{Metallicity Variation with Location} % (fold)
\begin{figure}
    \centering
    \includegraphics[width=0.45\textwidth]{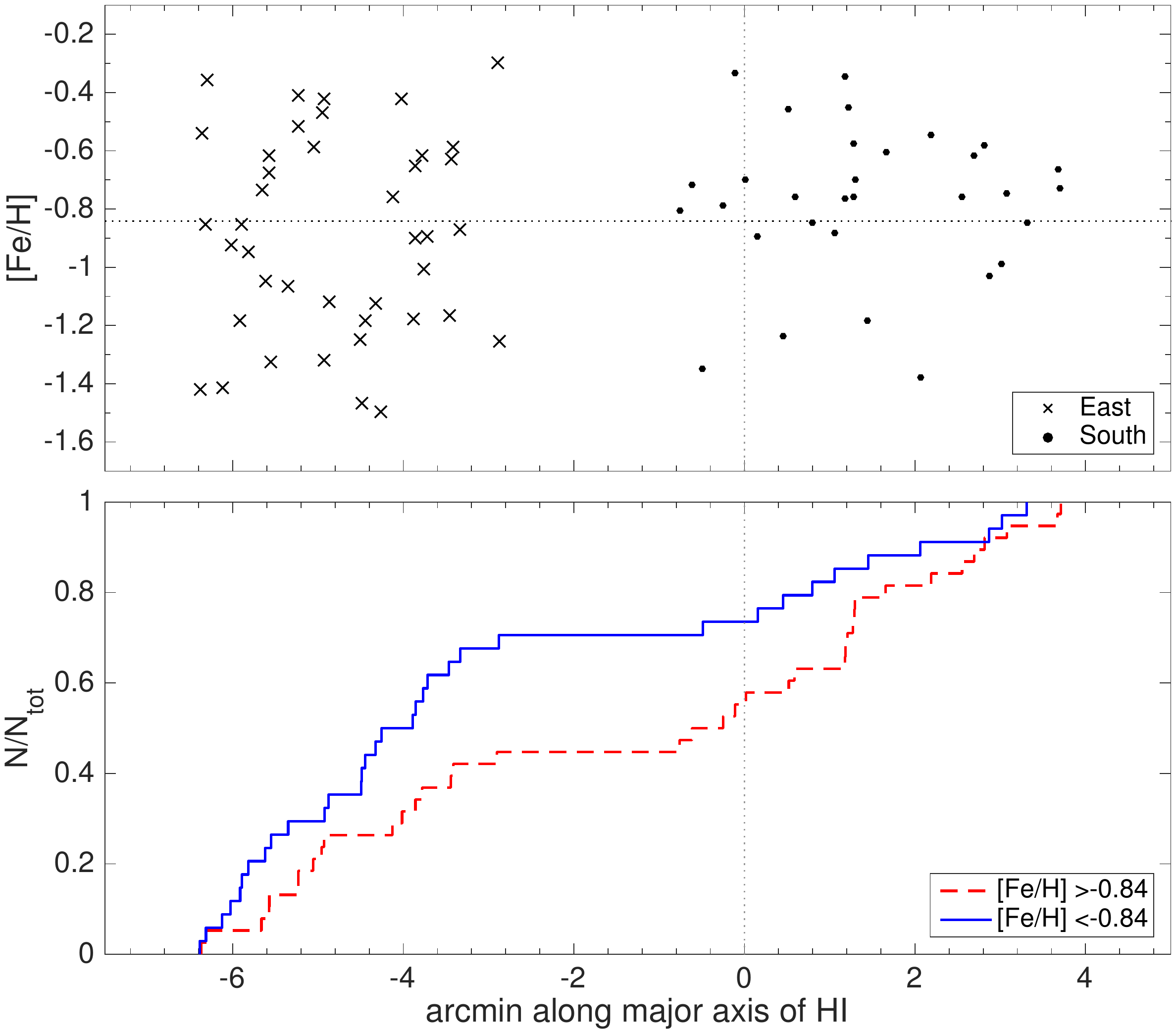}
    \caption{Using the centre of the \HI{} envelope as our origin, this shows the spatial variation of metallicity across our fields. The top plot is a simple scatter of our data with mean metallicity marked. The bottom plot segregates the data into metal rich and metal poor subsamples and displays CDFs for each of these.}
    \label{fig:Rad_met_HI}
\end{figure}

\label{sub:spatial_chemistry}
There is no clear-cut metallicity gradient in our data, due not only to the limited spatial extent of this survey, but also the moderate scatter. For example, a metallicity gradient as large as that of the SMC ($-0.075$ dex kpc\per{}; \citealt{Dobbie2014a}) would yield a difference in metallicity between the East field and the South field of $\approx$0.06 dex; the dispersion in our [Fe/H] alone is 0.31 dex which could easily disguise such a gradient in a small sample. The cumulative distribution functions on the other hand, allow us to search the sample for any tendency of high or low metallicity stars to cluster in or avoid certain places. Looking at the bottom panel of Figure \ref{fig:Rad_met_HI}, it can be seen that in the outer field ($-6\, '$\les{}$r_{ell}$\les{}$-5\, '$) that there are equal densities of metal poor and metal rich stars. For $-5\, '$\les{}$r_{ell}$\les{}$-3\, '$, the number of metal poor stars is increasing much more rapidly than metal rich stars, indicating a dominance of metal poor stars at intermediate radii. In the inner field ($-1\, '$\les{}$r_{ell}$\les{}$2\, '$) we see the opposite trend, with the number of metal rich stars increasing faster than metal poor.

These results are in agreement with results from \cite{Gallart1996c} and \cite{Cannon2012}. Whilst \cite{Gallart1996c} suggested the possibility of a metallicity gradient, \cite{Cannon2012} directly measured a concentration of young and intermediate age stars towards central regions, they also found that the older populations are more evenly distributed throughout NGC~6822. We see clear evidence to support this in our results, despite our lack of data at radii $\gtrsim$ 10$\, '$. So even without drastically offset target locations, the spatial metallicity variations are marked enough that they can be noted in our sample.

% subsection spatial_chemistry (end)

\subsection{Kinematic-Metallicity Correlation} % (fold)
\begin{figure}
    \centering
   \includegraphics[width=0.45\textwidth]{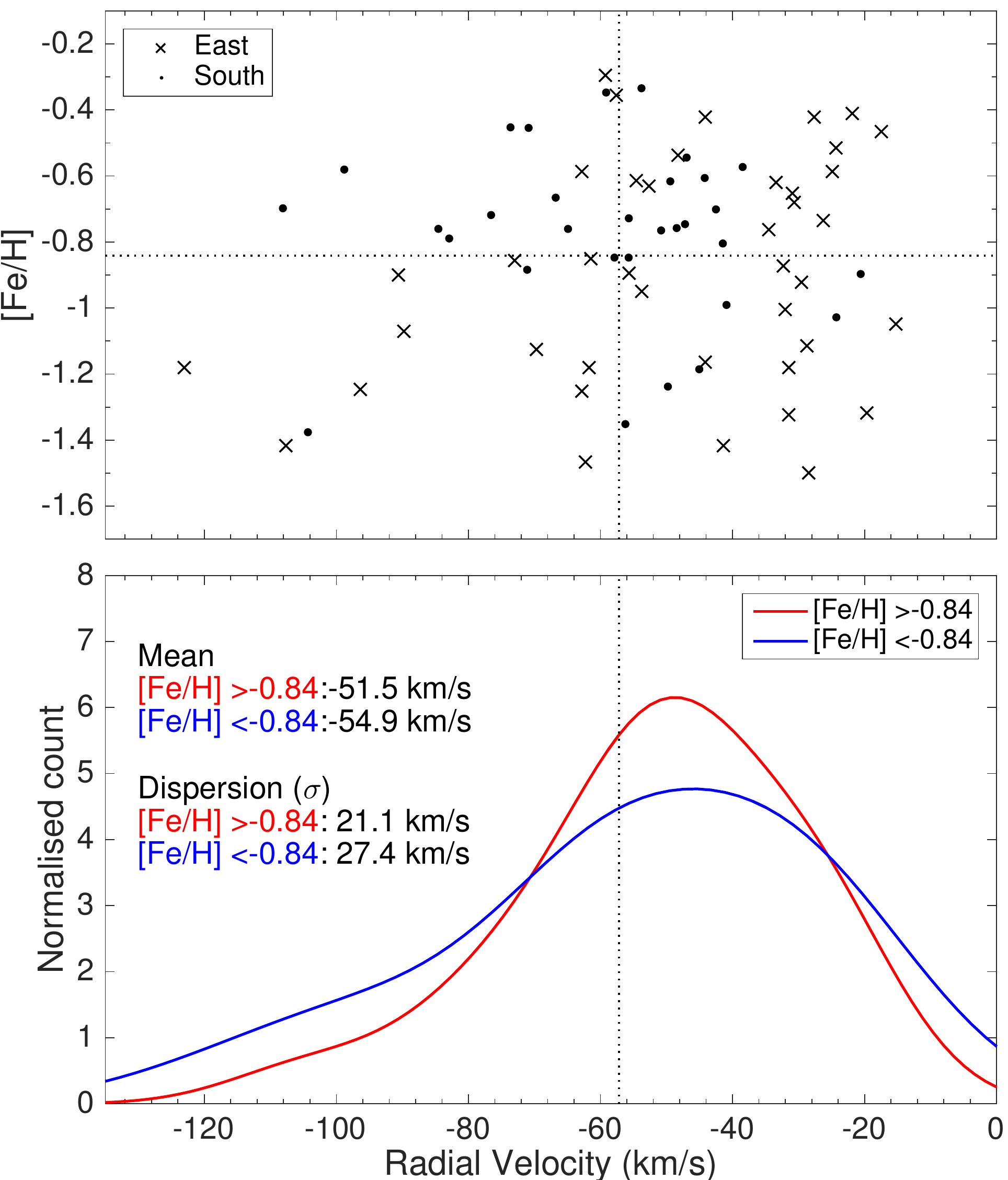}
    \caption{The top panel of this figure shows the distribution of stellar metallicity versus radial velocity, with means of both metallicity and radial velocity marked. The bottom panel is bifurcated into metal rich and metal poor star (objects above and below the mean metallicity) to compare the velocity distribution of these subsamples.}
    \label{fig:chemodynamics}
\end{figure}

\label{sub:dynamical_chemistry}
With stellar velocities and metallicities of more than 70 red giant stars in NGC~6822, we can attempt some simple chemodynamic analysis. Looking at the top panel in figure \ref{fig:chemodynamics} reveals that there is a tendency for the more metal rich stars to be clustered about the mean velocity of our targets. Comparing this to the metal poor stars they show no such tendency; this is consistent with dynamical heating of the older more metal poor populations over their lifetimes. 

Upon segregating the targets into two metallicity bins - one on either side of \mean{[Fe/H]} $=-0.84\,(\pm\,0.04)$ dex - and producing a PDF for each bin, two things become apparent. Firstly it can be seen that the mean velocity of the metal poor and metal rich populations are practically identical. Labelling these metal poor (\textsc{mp}) and metal rich (\textsc{mr}) we find $\langle v_{\rm MR}\rangle= -51.5 \pm 2.4$ km s\per{} and $\,\,\langle v_{\rm MR}\rangle = -54.9 \pm 2.3$ km s\per{}. This indicates that the bulk motion of each of these populations match well with the group mean. The velocity dispersions do on the other hand, vary significantly; here we find $\sigma_{v_{\rm MP}}=27.4$ km s\per{} and $\sigma_{v_{\rm MR}}=21.1$ km s\per{}. These chemodynamics are similar to those found in the Sculptor dSph by \cite{Tolstoy2001} in WLM dwarf irregular by \cite{Leaman2009}, and in the Magellanic Clouds (e.g., \citealt{Cole2005}, \citealt{Dobbie2014b}). However with our data a KS test returns a p-value of 0.60 indicating that the velocity distributions of these two sub-populations should not be considered independently, making it impossible to conclusively comment on any correlation between kinematics and metallicity. 

\section{Summary \& Conclusions}
The methods used in this paper allow for a more accurate metallicity measurement from a sample with relatively few targets. By using techniques for improved rejection of foreground dwarf contaminants, we have eliminated the tendency for their presence to bias the metallicity of NGC~6822 towards lower values due to their small CaT. Point to point reddening corrections have helped to alleviate some of the issues with measuring metallicity of more highly reddened regions allowing for a greater sensitivity to a metallicity gradient across variably reddened fields. Resulting metallicities are marginally higher than those of similar studies, but because of our corrections they likely represent the intermediate aged stellar population more accurately.

Our velocity measurements are consistent with other studies, and show that the older, more metal-poor population is dynamically { hotter} than its younger counterparts. { The} young, metal-rich stars tend to cluster towards central galactic radii indicating an outside-in formation process. This decrease in mean stellar age with decreasing radius was directly measured by Cannon et al.\ (2012), and is typically of dwarf irregular galaxies in the Local Group (e.g., Skillman et al.\ 2014).

Despite a similar sample size and higher precision in our velocity measurements  ($\pm$7 km s\per{} c.f. $\pm$15 km s\per{}) we see no evidence of a polar ring type velocity profile such as that suggested by \cite{Battinelli2006}. We have a much smaller sample size than \cite{Kirby2013}, whose spatial extent was largely along the position angle suggested for the polar ring by Battinelli et al.. Therefore we cannot add much more to the constraints already placed on this hypothesis by Kirby et al. However, our data extend further to the east than do Kirby's data, nearly parallel to the position angle of the HI rotation curve \citealt{Weldrake2003}. { The velocity difference between the two fields is in the same sense and of similar magnitude to that predicted if the stars and gas rotate in the same sense, but we cannot yet unambiguously ascribe the difference to rotation or derive a rotation curve of red giants in NGC~6822}. This may be due to the confounding effect of multiple stellar populations combined with dynamical heating over time and our relatively small sample size, or it could be attributable to the currently observed HI gas having a different dynamical origin than the gas from which the red giants formed, several Gyr ago. This can only occur if the gas has not had time to dynamically relax into the potential created by the stars and dark matter in the system. Further spectroscopic studies at a range of radial distances and position angles are required in order to fully diagnose the dynamical state of NGC~6822.
	
% subsection dynamical_chemistry (end)

\section*{Acknowledgements}
It is a pleasure to thank support astronomers Poshak Gandhi and Stefano Bagnuolo for their able assistance during observations. Travel support for AC was provided by the Nederlandse Onderzoekschool voor Astronomie (NOVA). { We would also like to thank the reviewer, Evan Kirby, for his knowledgeable and helpful feedback, and for providing us with extra information and new calculations relevant to his results, which significantly improved this paper. We similarly thank Erwin de Blok for providing the necessary data for plotting the H \textsc{i} contours in Figure \ref{fig:CCDregs}.}

\appendix
\section[]{Stellar data}
 
\begin{table*}
  \centering
  \caption[Known parameters of the 24 templates]{The parameters that are known for the 24 templates. Metallicities are identical for stars that were sampled from the same population. The star with no metallicity is the template dwarf star.}
    \begin{tabular}{ccccccccc}
    \hline
    \hline
    \textsc{star id} & \textbf{$v_{rad}$} & \textbf{W$_{8542}$} & \textbf{$\delta$} & \textbf{W$_{8662}$} & \textbf{$\delta$} & \textbf{[Fe/H]} & \textbf{W$_{Mg}$} & \textbf{$\delta$} \\
    \hline
    I260  & -103  & 2.14  & 0.01  & 1.72  & 0.01  & -2.07 & 0.25  & 0.02 \\
    I239  & -85.0 & 1.59  & 0.01  & 1.44  & 0.01  & -2.07 & 0.13  & 0.01 \\
    I119  & -100  & 1.63  & 0.01  & 1.42  & 0.01  & -2.07 & 0.25  & 0.01 \\
    AL12  & 166   & 2.19  & 0.01  & 1.68  & 0.01  & -1.69 & 0.23  & 0.01 \\
    AL15  & 166   & 1.99  & 0.01  & 1.57  & 0.01  & -1.69 & 0.23  & 0.01 \\
    6     & 205   & 2.21  & 0.02  & 1.86  & 0.02  & -1.37 & 0.26  & 0.02 \\
    15    & 207   & 3.09  & 0.01  & 2.35  & 0.01  & -1.37 & 0.45  & 0.01 \\
    241   & 218   & 2.41  & 0.01  & 2.00  & 0.01  & -1.37 & 0.36  & 0.01 \\
    237   & 217   & 2.69  & 0.01  & 2.16  & 0.01  & -1.37 & 0.35  & 0.01 \\
    89    & 204   & 2.49  & 0.01  & 1.93  & 0.02  & -1.37 & 0.54  & 0.01 \\
    209   & 213   & 2.51  & 0.02  & 2.09  & 0.01  & -1.37 & 0.44  & 0.02 \\
    22    & 87.0  & 2.82  & 0.02  & 2.14  & 0.02  & -0.47 & 0.59  & 0.01 \\
    27    & 15.0  & 2.79  & 0.02  & 2.19  & 0.05  & -0.47 & 0.89  & 0.01 \\
    3133  & 15.0  & 3.31  & 0.09  & 2.36  & 0.03  & -     & 0.74  & 0.07 \\
    4151  & 17.1  & 3.77  & 0.11  & 2.94  & 0.04  & -0.48 & 0.80  & 0.05 \\
    KR009 & 59.9  & 3.21  & 0.02  & 2.59  & 0.02  & -0.48 & 0.67  & 0.01 \\
    KR002 & 70.9  & 3.85  & 0.02  & 3.01  & 0.02  & -0.32 & 0.88  & 0.01 \\
    KR012 & 76.7  & 3.32  & 0.02  & 2.62  & 0.02  & -0.32 & 0.80  & 0.02 \\
    KR005 & 71.7  & 3.47  & 0.02  & 2.75  & 0.02  & -0.32 & 0.73  & 0.01 \\
    KR016 & 73.7  & 3.32  & 0.02  & 2.61  & 0.02  & -0.32 & 0.73  & 0.01 \\
    KR003 & 87.6  & 3.68  & 0.02  & 2.88  & 0.02  & -0.32 & 0.78  & 0.01 \\
    KR028 & 77.8  & 2.87  & 0.02  & 2.49  & 0.02  & -0.32 & 0.63  & 0.01 \\
    \hline
    \end{tabular}%
  \label{tab:templates}%
\end{table*}%
\newpage

\begin{sidewaystable*}
\small
  \caption[Amalgamation of important parameters for our 72 NGC~6822 red giant branch stars]{Amalgamation of important parameters for our 72 NGC~6822 red giant branch stars. Carbon stars are indicated with a subscript $C$, the three south field stars for which the $V$ magnitudes were estimated are marked with a dagger. The error in the magnesium line width is omitted, but is \apr{}0.1 for most cases. Units are omitted to allow the tables to fit on the pages. Units are: degrees, magnitude, km s\per{}, \AA{}, dex; where appropriate. Each $\delta$ represents the \textit{random} error of the previous column, for systematic errors see Table \ref{tab:errors}.}
    \begin{tabular}{cccccccccccccccc}
    \hline
    \multicolumn{1}{c}{\textsc{star id}} & \textbf{R.A.} & \textbf{Decl.} & \textbf{$V$} & \textbf{$\delta$} & \textbf{$v_{rad}$} & \textbf{$\delta$} & \textbf{$I_{\rm TRGB}$} & \textbf{$\delta$} & \textbf{$\Sigma$W$_{\rm Ca}$} & \textbf{$\delta$} & \textbf{$V_{\rm HB}$} & \textbf{$\delta$} & \textbf{[Fe/H]} & \textbf{$\delta$} & \textbf{W$_{Mg}$}  \\
    \hline
    \hline
    HS02  & 296.374 & -14.886 & 22.15 & 0.01  & -32.3 & 1.3   & 20.17 & 0.45  & 7.86  & 0.23  & 24.99 & 0.46  & -0.87 & 0.08  & 0.64 \\
    HS03  & 296.431 & -14.884 & 22.31 & 0.04  & -30.6 & 1.3   & 20.18 & 0.33  & 8.28  & 0.39  & 25.00 & 0.34  & -0.68 & 0.09  & 0.78 \\
    HS04  & 296.382 & -14.881 & 21.69 & 0.02  & -55.8 & 1.3   & 20.17 & 0.33  & 8.14  & 0.19  & 24.99 & 0.34  & -0.89 & 0.04  & 0.82 \\
    HS05  & 296.420 & -14.878 & 21.91 & 0.02  & -24.4 & 1.2   & 20.16 & 0.14  & 9.01  & 0.20  & 24.98 & 0.17  & -0.52 & 0.02  & 0.79 \\
    HS06  & 296.381 & -14.876 & 22.02 & 0.03  & -54.7 & 1.3   & 20.17 & 0.06  & 8.66  & 0.17  & 24.99 & 0.10  & -0.61 & 0.01  & 0.96 \\
    HS07  & 296.441 & -14.874 & 22.05 & 0.03  & -41.3 & 1.2   & 20.16 & 0.11  & 6.43  & 0.51  & 24.98 & 0.14  & -1.42 & 0.10  & 0.70 \\
    HS08  & 296.382 & -14.872 & 22.16 & 0.03  & -31.0 & 1.3   & 20.16 & 0.31  & 8.45  & 0.18  & 24.98 & 0.32  & -0.65 & 0.04  & 0.92 \\
    HS09  & 296.426 & -14.869 & 21.94 & 0.02  & -15.3 & 1.3   & 20.15 & 0.14  & 7.51  & 0.28  & 24.98 & 0.16  & -1.05 & 0.04  & 0.69 \\
    HS10  & 296.415 & -14.867 & 22.18 & 0.03  & -21.7 & 1.3   & 20.14 & 0.05  & 9.09  & 0.22  & 24.97 & 0.10  & -0.41 & 0.02  & 0.54 \\
    HS11  & 296.379 & -14.864 & 21.60 & 0.02  & -31.6 & 1.3   & 20.12 & 0.26  & 7.39  & 0.23  & 24.95 & 0.27  & -1.18 & 0.04  & 0.54 \\
    HS12  & 296.352 & -14.862 & 21.82 & 0.02  & -62.9 & 1.3   & 20.12 & 0.40  & 7.02  & 0.31  & 24.95 & 0.41  & -1.25 & 0.08  & 0.48 \\
    HS13  & 296.387 & -14.860 & 21.97 & 0.02  & -28.4 & 1.3   & 20.13 & 0.04  & 6.24  & 0.19  & 24.96 & 0.10  & -1.50 & 0.02  & 0.62 \\
    HS14  & 296.405 & -14.857 & 22.05 & 0.03  & -17.6 & 1.3   & 20.13 & 0.22  & 9.03  & 0.17  & 24.96 & 0.23  & -0.47 & 0.03  & 0.68 \\
    HS15  & 296.364 & -14.855 & 22.06 & 0.03  & -44.1 & 1.3   & 20.13 & 0.05  & 7.10  & 0.25  & 24.96 & 0.10  & -1.16 & 0.03  & 0.64 \\
    \ldots  & \ldots & \ldots & \ldots & \ldots & \ldots & \ldots   & \ldots & \ldots  & \ldots  &\ldots  &\ldots & \ldots  & \ldots & \ldots  & \ldots \\
  \hline
    \end{tabular}%
  \label{tab:bigtableofstardata}%
\end{sidewaystable*}%
\bsp

\label{lastpage}
\end{document}